\documentclass[preprint]{aastex}
\usepackage{graphicx,amsmath}
\usepackage[normalem]{ulem}
\usepackage{alltt}
\usepackage{placeins}
\usepackage{multirow,rotating}
\usepackage{datetime}

\usepackage{multicol}
\usepackage{afterpage}
\usepackage{color}
\usepackage{capt-of}
\newcommand{\hi}{\ion{H}{1}}   
\newcommand{\Ha}{H$\alpha$}

\shortauthors{Moorman et al.}
\shorttitle{SSFRs \& SFEs of Void Dwarf Galaxy}
\begin{document}

\title{On the Star Formation Properties of Void Galaxies }  

\author{Crystal M. Moorman\altaffilmark{1}, Jackeline Moreno, Amanda White\altaffilmark{2}, Michael S. Vogeley} 
\affil{Department of Physics, Drexel University, 3141 Chestnut Street, Philadelphia, PA 19104} 
\email{crystal.m.moorman@drexel.edu} 
\altaffiltext{1}{Physics Department, Lynchburg College, 1501 Lakeside Drive, Lynchburg, VA 24551} 
\altaffiltext{2}{American Museum of Natural History, Central Park West \& 79th St, New York, NY 10024} 
\author{Fiona Hoyle} 
\affil{Pontifica Universidad Catolica de Ecuador, 12 de Octubre 1076 y Roca, Quito, Ecuador} 
\and 
\author{Riccardo Giovanelli, Martha P. Haynes} 
\affil{Center for Radiophysics and Space Research, Space Sciences Building, Cornell University Ithaca, NY 14853} 

\begin{abstract}

We measure the star formation properties of two large samples of galaxies from the SDSS in large-scale cosmic voids 
on time scales of 10 Myr and 100 Myr, using \Ha\ emission line strengths and GALEX $FUV$ fluxes, respectively. 
The first sample consists of 109,818 optically selected galaxies. 
We find that void galaxies in this sample have higher specific star formation rates (SSFRs; star formation rates per unit stellar mass) 
than similar stellar mass galaxies in denser regions. 
The second sample is a subset of the optically selected sample 
containing 8070 galaxies with reliable \hi\ detections from ALFALFA. 
For the full \hi\ detected sample, SSFRs do not vary systematically with 
large-scale environment. 
However, investigating only the \hi\ detected dwarf galaxies reveals a trend towards higher SSFRs in voids. 
Furthermore, we estimate the star formation rate per unit \hi\ mass (known as the star formation efficiency; SFE) of a galaxy, 
as a function of environment. 
For the overall \hi\ detected population, we notice no environmental dependence. 
Limiting the sample to dwarf galaxies still does not reveal a statistically significant difference between SFEs in voids versus walls. 
These results suggest that void environments, on average, provide a nurturing environment for dwarf galaxy evolution  
allowing for higher specific star formation rates while forming stars with similar efficiencies to those in walls. 
\end{abstract}


\section{INTRODUCTION}
\label{sec:sfr_intro}

Large-scale cosmic voids make up about 60\% of the volume of our Universe \citep{GellerHuchra1989,daCosta1988,Pan2012}. 
While not completely empty, these voids are underdense enough that 
gas-stripping galaxy interactions are exceptionally rare, making these voids a 
unique environment for studying the formation and evolution of galaxies. 
According to $\Lambda$ Cold Dark Matter ($\Lambda$CDM) simulations (e.g. \cite{Hoeft2006}), 
voids should have a density of low-mass halos of about 1/10th the cosmic mean. 
\cite{Karachentsev2004} find that the measured density of 
observed bright galaxies in the Local Void closely matches this prediction, 
but the density of dwarf galaxies in the Local Void is measured to be 1/100th that of the cosmic mean. 

A posible explanation for this lack of low-mass, faint galaxies is that void dwarf galaxies may have their star formation suppressed \citep[e.g.][]{Koposov2009}. 
One critical test of comparison between simulations and observations is through the 
relation of the predicted dark matter halo mass function and the observed galaxy luminosity and/or mass functions. 
Both simulations and semi-analytic models apply methods of star formation suppression to simulated dwarf ``galaxies'' 
in attempt to more closely match observations. 
Semi-analytic models (e.g. \cite{Benson2002}) suggest that reionization could suppress the dwarf galaxies. 
On the other hand, \cite{Hoeft2006} find in hydrodynamic simulations that 
adding in effects of UV photo-heating before and after reionization is not enough to 
match the slope of the observed mass function. 
Simulations more targeted at determining effects of large-scale environment  on galaxies include the 
high resolution hydrodynamical simulations of void and cluster environments by \cite{Cen2011}. 
This adaptive mesh refinement simulation is the strongest predictor of void dwarf galaxy properties to date. 
The simulation predicts void dwarf galaxies at $z=0$ will have high specific star formation rates. 
It predicts that these void dwarf galaxies are able to continue forming stars, because the entropy of the gas in voids 
is below the threshold at which the galaxies' cooling times exceeds a Hubble time. 

Previous studies based on galaxy surveys suggest that galaxies within 
voids have higher star formation rates per stellar mass 
(specific star formation rate; SSFR) than galaxies in walls.  
\cite{Rojas2005} use a sample of $\sim1000$ underdense SDSS galaxies, 
corresponding to $\delta\rho/\rho<-0.6$, and find that void galaxies have 
higher SSFRs given their stellar mass. 
\cite{vonBenda-Beckmann2008} find stronger star formation suppression in the field than in voids 
using a sample of faint galaxies from the 2dFGRS. 
These authors also find that galaxies towards void centers have 
a weak tendency to have higher rates of star formation.   
In contrast, using a collection of 60 targeted \hi\ imaged galaxies from the 
Void Galaxy Survey \citep[VGS;][]{Stanonik2009, Kreckel2011, Kreckel2012}, 
\cite{Kreckel2012} find SSFRs of their void galaxies 
to be similar to SSFRs of galaxies from GALEX Arecibo SDSS Survey \citep[GASS;][]{Catinella2010} in 
average environments. It is important to note that \hi\ selection has a strong effect on the sample.
\cite{Ricciardelli2014} compare optically selected galaxies from within the 
``maximal spheres'' of voids to a sample of ``shell'' galaxies and find 
no environmental dependence on SSFR. 
Environmental-based results from any work are certainly sensitive to the author's definition of environment. 
\cite{Icke1984} and \cite{vandeWeygaert1993} show that voids 
evolve to become more spherical with time; 
however, current measurements of actual voids from \cite{PanPhD2011} show that 
voids are more ellipsoidal with a tendency to be prolate. 
The use of spheres and spherical shells to define environment in \cite{Ricciardelli2014} results in the 
``shell'' sample potentially being contaminated by void galaxies. 
It is, therefore, unsurprising that \cite{Ricciardelli2014} find no environmental dependence on SSFR. 
We discuss these results in more detail in Section~\ref{sec:compare}.

Overall, it seems that void galaxies tend to have higher SSFRs than 
galaxies in average density regions \citep{Rojas2005,vonBenda-Beckmann2008}. 
Thus, we wonder if there is a trend in how efficiently galaxies are 
converting their gas into stars across environments, allowing void dwarf galaxies to continue 
their star formation at late times. 
The efficiency with which a galaxy transforms its gas into stars is called 
Star Formation Efficiency (SFE) and is defined as the 
SFR normalized by the \hi\ mass of the galaxy (SFE=SFR/$M_{HI}$). 
To determine how effectively galaxies are forming stars from their gas, 
we require an \hi\ mass for each galaxy. 
This requirement will impose an \hi\ selection bias on our sample, so we 
must carefully examine what effect \hi\ selection has on SFRs, before making 
environmental comparisons. 

\hi\ surveys typically detect blue, gaseous, actively star forming galaxies.  
\cite{Huang2012} find that selecting only \hi\ detections results in 
overall higher SSFRs than an optically selected sample. 
This is due primarily to the removal of most passive (inactive) galaxies 
from the optically selected sample.  \cite{Kreckel2012} study the efficiency of 
galaxies in the VGS and find two galaxies with which they can compare to 
similar stellar mass galaxies in denser regions. 
These two galaxies happen to have higher SFEs than the GASS galaxies to which 
they compare their observations. This hints that void galaxies \emph{may} have 
higher SFEs than galaxies in average environments, but 
strong conclusions cannot be made with a sample of only two galaxies. 

In this paper, we present the environmental effects on the  
specific star formation rate and the star formation efficiency of dwarf galaxies in voids 
using optical data from the SDSS DR8, \hi\ data from the 
ALFALFA Survey, and UV data from GALEX. 
For the first time, we determine the large-scale environmental impact on the star formation properties of 
dwarf galaxies down to $M_r=-13$. 
Throughout this work, we assume $\Omega_m=0.26$ 
and $\Omega_{\Lambda}=0.74$ when calculating comoving coordinates.

\section{NASA-Sloan Atlas}
\label{sec:nsa_Data}
The parent data set that we use in this work is the NASA-Sloan Atlas (NSA) version 
\citep{Blanton2011}. 
The NSA is a collection of galaxies in the local Universe ($z\le0.055$) 
based primarily on the Sloan Digital Sky Survey Data Release 8 (SDSS DR8) spectroscopic catalog \citep{York2000,Aihara2011} 
and contains about 140,000 galaxies within the footprint of SDSS DR8. 
The catalog contains galaxy parameters and images from a combination of several catalogs across multiple wavelengths: 
SDSS DR8, NASA Extragalactic Database, Six-degree Field Galaxy Redshift Survey, 
Two-degree Field Galaxy Redshift Survey, ZCAT, WISE, 2MASS, GALEX, and ALFALFA. 
The NSA catalog re-analyzes the SDSS photometry 
in $u,g,r,i$, and $z$ bands using the background subtraction methods described in \cite{Blanton2011}. 
The photometry is redone to help eliminate the contamination of a dwarf galaxy sample by 
shredded or deblended galaxies 
as well as to improve background subtraction procedures for large galaxies. 
The spectroscopic data is also re-analyzed using methods of \cite{Yan2011} and \cite{Yan2012}. 
Distances within this catalog are obtained using the local velocity flow model of \cite{Willick1997}. 
Stellar masses in the catalog were estimated from K-correction fits \citep{Blanton_kcorrect}. 

\subsection{SDSS DR8}
\label{subsec:sdss_data} 
To determine SFRs on time scales of 10 Myr, 
we use the strength of the \Ha\ emission line, which we 
obtain from the SDSS DR8 parameters provided within the NSA. 
The SDSS is a wide-field multi-band imaging and spectroscopic survey that uses drift scanning 
to map about a quarter of the northern sky. SDSS employs the
2.5m telescope at Apache Point Observatory in New Mexico,
allowing it to cover $\sim10^4$ deg$^2$ of the northern hemisphere in the five band SDSS 
system--$u,g,r,i$, and $z$ \citep{Fukugita1996, Gunn1998}.  
Galaxies with Petrosian $r$-band magnitude $r < 17.77$ are selected for 
spectroscopic follow up \citep{Lupton2001,Strauss2002}. 
SDSS spectra are taken
using two double fiber-fed spectrographs and fiber plug plates covering
a portion of the sky $1.49^{\circ}$ in radius with a minimum fiber 
separation of 55 arcseconds \citep{Blanton2003SpectraSDSS}. 
The SDSS DR8 spectra used in the NSA were reduced using the Princeton 
spectroscopic reduction pipeline.  
Flux measurements for this catalog were estimated using the code from \cite{Yan2011} 
that calibrates the flux by matching the synthetic $r$-band magnitude to the apparent $r$-band fiber magnitude 
and subtracts the stellar continuum modeled in \cite{Bruzual2003}. 
These methods are detailed in \cite{Yan2011} and \cite{Yan2012}. 

\subsection{GALEX}
\label{subsec:galex_data} 
To determine SFRs on timescales of 100 Myr, we need photometric information 
in the far and near ultraviolet bands ($FUV$ and $NUV$, respectively). 
To obtain this information, we utilize data from the Galaxy Evolution Explorer (GALEX) \citep{GALEX2005}. 
GALEX is an 0.5m orbiting space telescope that images the sky in $FUV$ and $NUV$ photometric bands 
across 10 billion years of cosmic history. Portions of the GALEX GR6 footprint overlap with the northern sky 
surveyed in SDSS DR8. In this work, we use the cross-matched NSA-GALEX GR6 catalog 
provided by the NSA team to obtain all UV parameters. 

\subsection{ALFALFA}
\label{subsec:alfalfa_data} 
To determine whether galaxies in voids are more efficient at forming 
stars than galaxies of similar mass in denser regions, we need an estimate of each galaxy's \hi\ mass.  
We utilize the Arecibo Legacy Fast ALFA (ALFALFA) Survey \citep{ALFALFAII,ALFALFAI} 
to obtain \hi\ information for our galaxies. 
ALFALFA is a large-area, blind extragalactic \hi\ survey that will detect over 30,000 sources out to $z\sim0.06$ 
with a median redshift of $z\sim0.027$, over 7000 deg$^2$ of sky upon completion.
ALFALFA has a 5$\sigma$ detection limit of 0.72 Jy km s$^{-1}$ for a source with a profile 
width of 200 km s$^{-1}$ \citep{ALFALFAII} allowing for the detection of sources 
with \hi\ mass down to $M_{HI}=10^8 M_{\odot}$ out to 40 Mpc. 
At a redshift of $z\sim0.02$, the distance out to which we can observe dwarf galaxies, ALFALFA allows for the detection of  \hi\ masses down to $M_{HI}=10^{8.63} M_{\odot}$ for sources with a profile width of 200 km s$^{-1}$. 

We use the most recent release of the ALFALFA Survey, $\alpha.40$ \citep{Haynes2011}, 
which covers $\sim$2800 deg$^2$. 
For this work, we are interested in sources lying in the Northern Galactic Hemisphere within $z=0.05$. 
The $\alpha.40$ catalog overlaps this region in two strips in the R.A. range $07^h30^m<$R.A.$<16^h30^m$: 
$04^{\circ}<$ Dec$<16^{\circ}$  and  $24^{\circ}<$ Dec $<28^{\circ}$.
This region contains 8,070 \hi\ detections within this volume. 
Each detection in the catalog is flagged as either Code 1, 2, or 9. 
Code 1 objects are reliable detections with S/N $>6.5$; Code 2 objects have S/N$<4.5$, but coincide with optical counterparts with known 
redshift similar to \hi\ detection redshift; and Code 9 objects correspond to high velocity clouds. 
An ALFALFA team member checked the cross-matching between ALFALFA and SDSS DR7 for each \hi\ source using information about each galaxy's color, morphology, redshift, etc. 
In this work, we only consider Code 1 and Code 2 detections. 
We use the \hi\ masses from this catalog (which utilize the heliocentric distances 
provided in the catalog assuming $h=0.7$) 
and use the NSA catalog for estimates of all other properties such as position, distance, and color. 
Because a blind \hi\ survey is biased towards detecting blue, faint, gaseous galaxies and 
because the survey volume covered by ALFALFA is substantially smaller than 
the full NSA survey volume, 
only a small fraction of galaxies in the NSA catalog have \hi\ information. 
Thus, we will separately track the effects of using an \hi\ selected sample 
starting in Section \ref{sec:methII}.

\section{DETERMINING ENVIRONMENT}
\label{sec:methI}
\subsection{Creating the Void and Wall Samples} 
\label{subsec:LSS}
We identify the large-scale environment of our sources 
by comparing the comoving coordinates of each galaxy to 
the void catalog of \cite{Pan2012}. 
This void catalog uses VoidFinder, the galaxy-based 
void finding algorithm of \cite{HoyleVogeley2002} (also see \citealt{ElAdPiran1997}). 
The algorithm grows spheres in the most underdense regions of a volume limited distribution of galaxies. 
Each sphere must live fully within the SDSS DR7 survey mask and must have radius $R\ge10$h$^{-1}$Mpc. 
If the volume of any two given spheres overlap by at least 10 percent, 
we say that the two spheres are associated with the same void. 
By merging the individual spheres, we allow VoidFinder to identify non-spherical voids. 
When physically comparing the location of our galaxies to the identified voids, 
we exclude galaxies living along the edge of the survey, because we cannot determine the true large-scale structure of 
galaxies on the survey boundaries 
(see \citealt{Pan2012}, \citealt{Hoyle2012}, or \citealt{Moorman2014} for a more detailed explanation). 
Galaxies identified as residing in a void 
are called void galaxies; those outside of voids are deemed wall galaxies. 
From the galaxies within $z<0.05$ with \Ha\ line fluzes in the NSA catalog, we identify 34,548 (31\%) void galaxies, 
70,950 (65\%) wall galaxies, and 4,320 (4\%) galaxies lying on the edge of the survey mask 
(unclassifiable as void or wall). 

\begin{figure}[ht]
 \centering 
  \includegraphics[scale=0.4,trim = 0.cm 0.cm 0.8cm 1.05cm, clip=True]{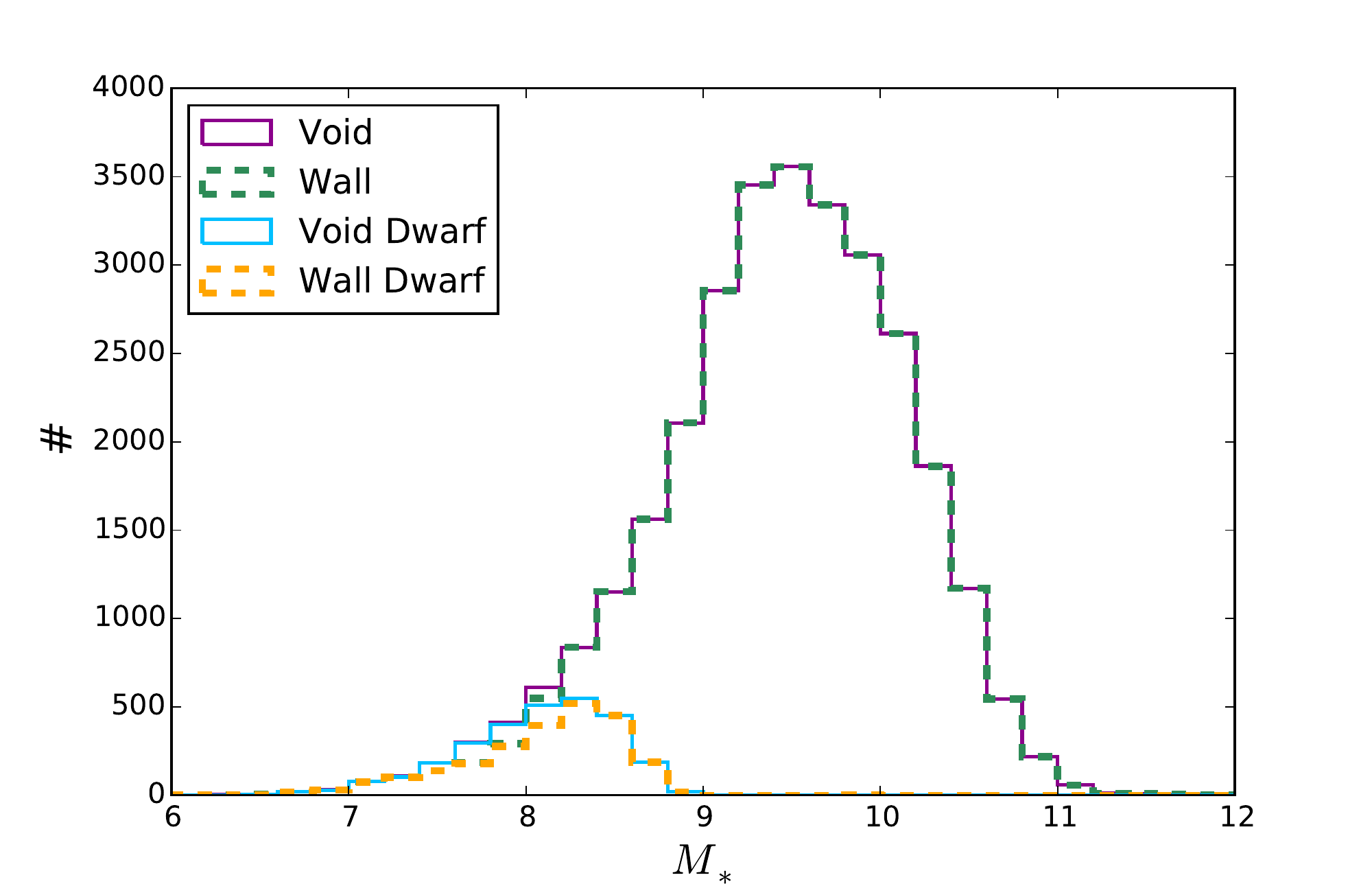} 
  \caption[Stellar mass distribution after mass matching]{
  Distribution of stellar masses of the void and wall galaxies in the NSA full and dwarf samples after 
  sparse sampling. 
  The wall galaxies are sparse-sampled so that the stellar mass distribution matches that of the voids.  
  Because the void stellar mass distribution is naturally shifted towards lower masses, 
  we find fewer wall galaxies than void galaxies at the low-mass end.  Thus the void and wall stellar 
  mass distributions are not identical at the low-mass end.}  
\label{fig:sm_distn}
\end{figure}
\begin{figure}[ht]
 \centering 
  \includegraphics[scale=0.4,trim = 0.cm 0.cm 0.8cm 1.05cm, clip=True]{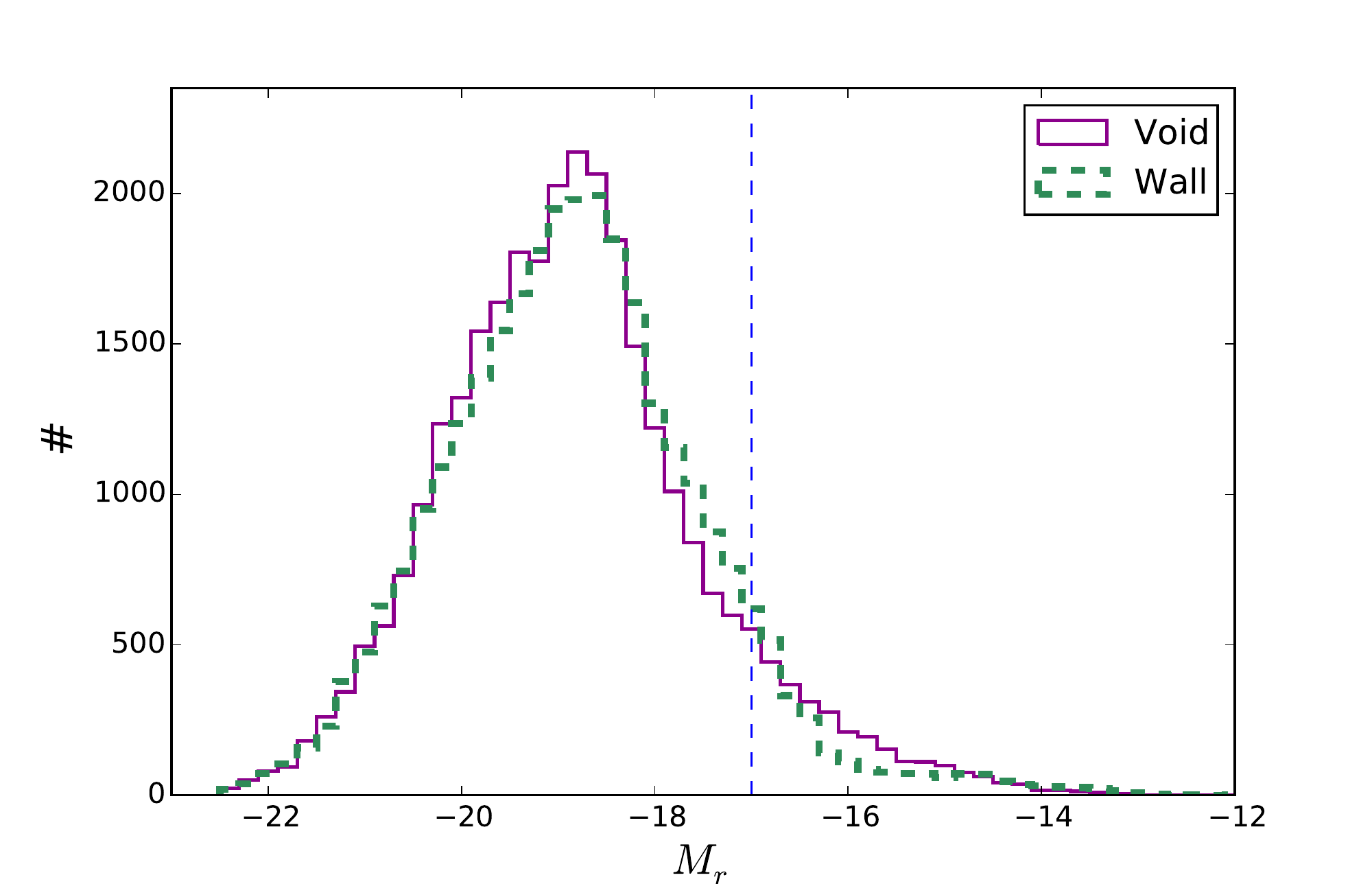}
  \includegraphics[scale=0.4,trim = 0.cm 0.cm 0.8cm 1.05cm, clip=True]{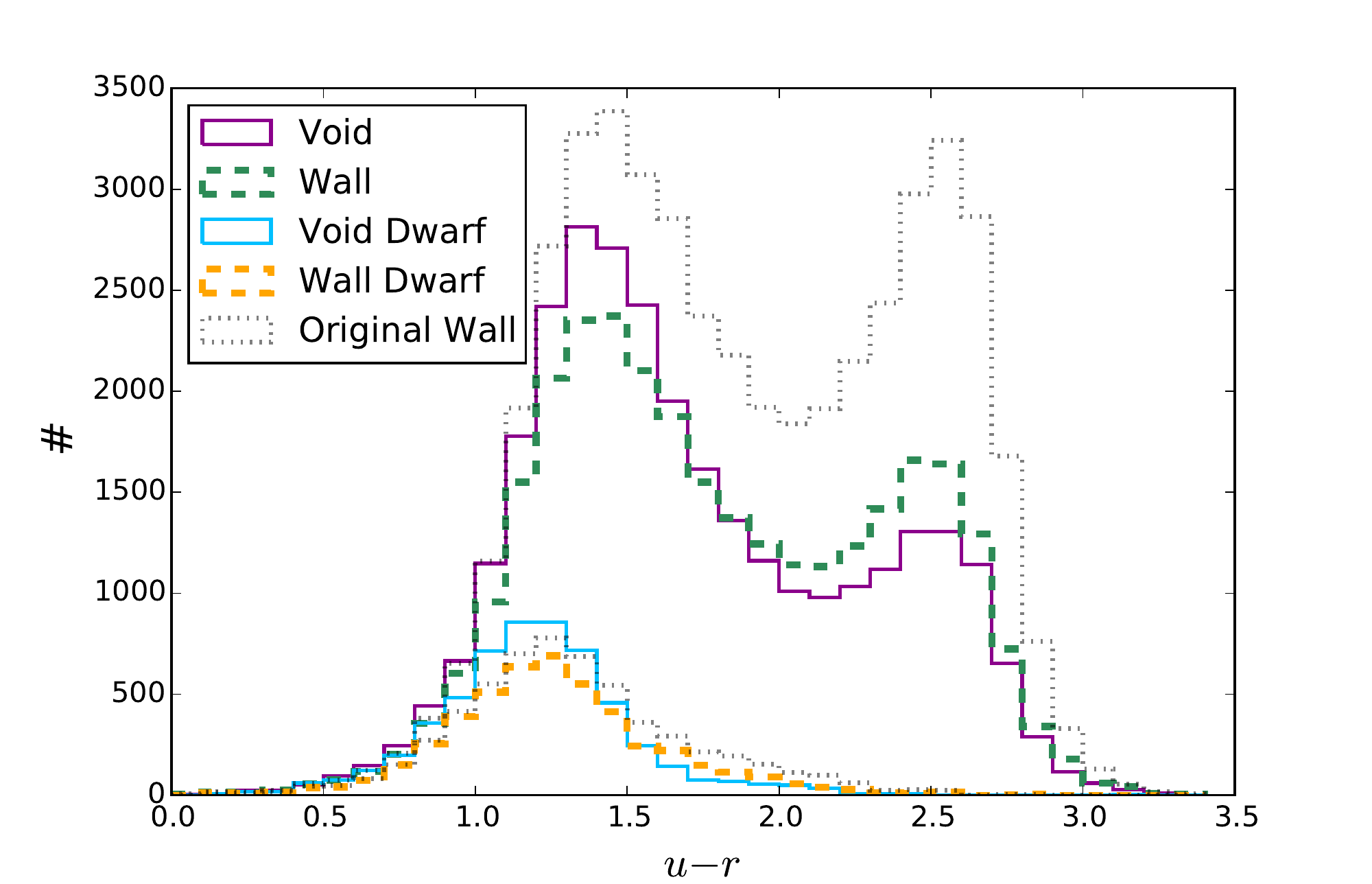}
 \caption[Magnitude and color distributions after mass matching]{Left: Distribution of absolute magnitude $M_r$ of 
 void and wall samples that have the same stellar mass distribution. 
 The bias towards brighter galaxies in walls is largely removed. 
 The blue dashed line indicates the division between dwarf galaxies and brighter galaxies. 
 Right: Distribution of $u-r$ color of void and wall samples that have the same stellar mass distribution. 
 The number of red galaxies within the walls is reduced, 
 but there is still a higher ratio of red to blue galaxies in walls than voids. 
 For clarity, the dwarf distributions have been amplified by a factor of 2.} 
\label{fig:sm_distn_mr_ur}
\end{figure}

Galaxies in voids are fainter and less massive than galaxies in denser regions. 
The characteristic luminosity of the void luminosity function shifts toward fainter 
luminosities by a factor of $\sim2.5$ (\citealt{Hoyle2005} and \citealt{Moorman2015}), 
and the characteristic \hi\ mass shifts toward lower \hi\ masses in the 
void \hi\ mass function \citep{Moorman2014}.  
Additionally, \cite{Goldberg2004} predict that the Dark Matter Halo mass functions 
shift toward lower masses in voids 
and measure this shift in \cite{Goldberg2005}. 
This dependence of the mass/luminosity function on large-scale environment 
is now firmly established. 
\cite{Alpaslan2015} have shown that stellar mass is a strong predictor of galaxy properties. 
It is well established that voids host fainter, less massive galaxies. In this analysis, we would like 
to determine if environment has an effect on the SFE of galaxies. 
To test for effects of large-scale environment beyond 
the differences in stellar mass distributions between voids and walls,  
we randomly sample galaxies from the wall distribution such that the 
new stellar mass distribution of the wall galaxies matches that of the void galaxies. 
See Figure \ref{fig:sm_distn} for the stellar mass-matched distributions of the 
full NSA sample and the dwarf-galaxy-only sample. 
As mentioned above, void galaxies are typically less massive than wall galaxies. 
Therefore, in some of the lowest mass bins, we have more void galaxies than wall galaxies. 
When we downsample the wall distribution, the lack of a sufficient sample of wall galaxies in these bins  
causes a very minor mismatch in the stellar mass distribution at masses lower than $\log(M_*)\sim8$. 
Downsampling the wall galaxy sample to match the stellar mass distribution of void galaxies 
also removes the bias towards brighter galaxies 
in the walls as seen in the left panel of Figure \ref{fig:sm_distn_mr_ur}.  
In the right panel of Figure~\ref{fig:sm_distn_mr_ur}, we see the effects of down-sampling the walls on $u-r$ color. 
Compared to the original optical color distributions seen in the walls 
(scaled by 0.75 and depicted as a gray dotted line in the figure), 
we see a decreased ratio of red to blue galaxies in walls, although not as low as that of the void sample. 
Similarly, stellar mass matching the wall dwarf distribution to match the 
void dwarf stellar mass distribution decreases the red galaxy population in the wall dwarf sample. 
(For wall dwarfs, we provide the original wall dwarf distribution as a scaled gray dotted line in the figure.) 

Of the galaxies with \hi\ information, we find 
2,737 ($34\%$) void galaxies and 5,154 ($64\%$) wall galaxies within $z<0.05$. 
The remaining (2\%) lie along the survey edges and are excluded due to the 
potential misclassification issues. 
We apply the same stellar mass-matched requirements to the \hi\ sample as discussed earlier. 
The stellar mass matching has similar effects on the \hi\ sample distributions. 
The $M_r$ distributions of void and wall samples become more similar. 
The $u-r$ distribution of the wall galaxies becomes less red overall. 
By its nature, the \hi\ sample primarily selects primarily blue, gas-rich galaxies \citep{Toribio2011Control,Huang2012,Moorman2015}. 
Because \hi\ selection preferentially selects gas-rich galaxies, 
we expect the wall galaxies in this sample to be primarily late-type galaxies, 
more similar to void galaxies than the full population of optically selected, wall galaxies. 
Additionally, the stellar mass matching accounts for the well-known shift towards less massive galaxies in voids;  
therefore, we do not expect that we will see much of a difference between the star formation properties 
of the \hi\ selected void and \hi\ selected wall galaxies.

\subsection{Creating the Small-Scale Density Samples}
\label{subsec:SSS}
\begin{figure}[ht]
\centering
  \begin{tabular}{@{}cc@{}}  
    \includegraphics[scale=0.415,trim = 0.4cm 0.2cm 1.65cm 1.05cm, clip=True]{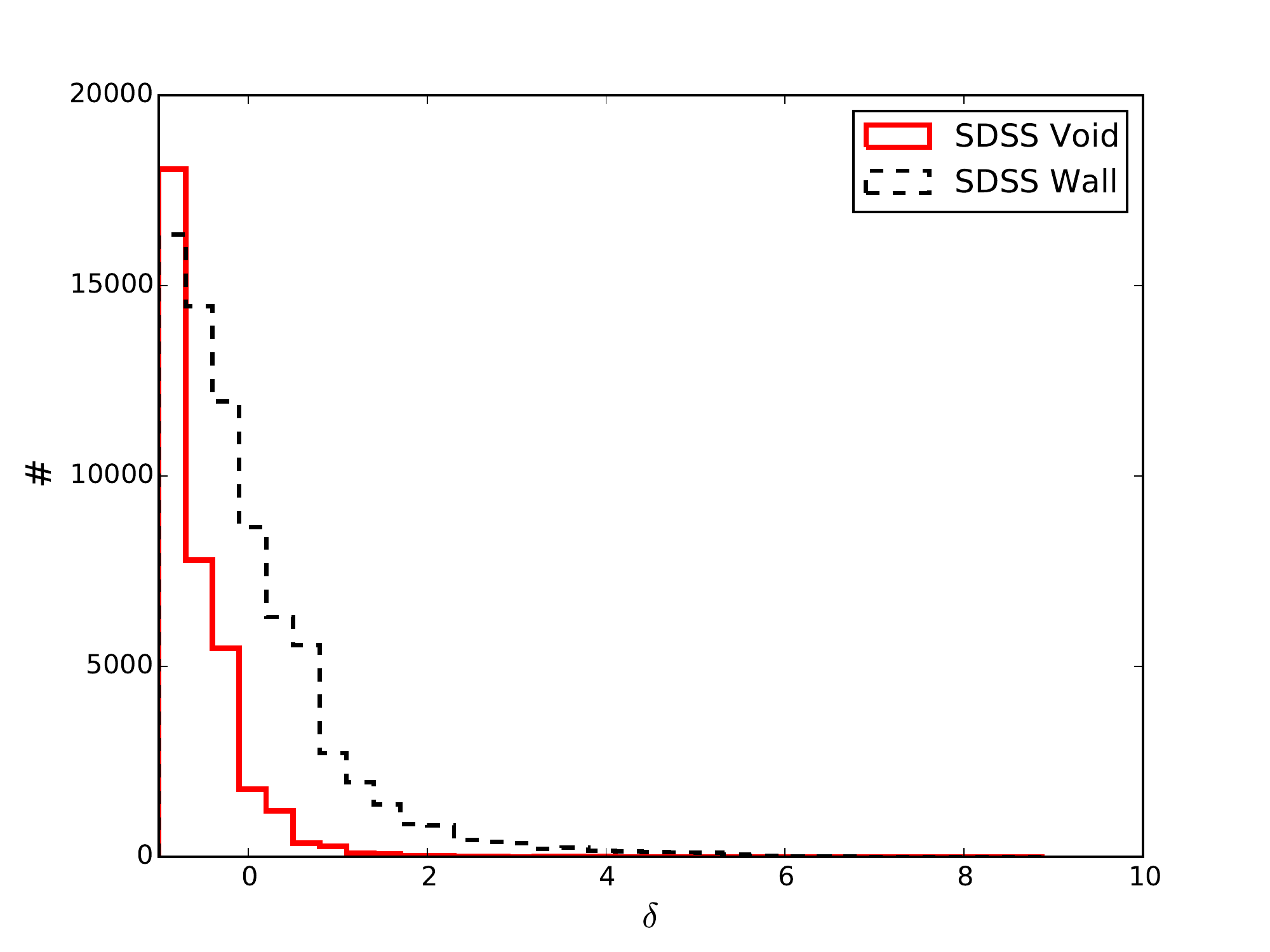} & 
    \includegraphics[scale=0.415,trim = 0.65cm 0.2cm 1.65cm 1.05cm, clip=True]{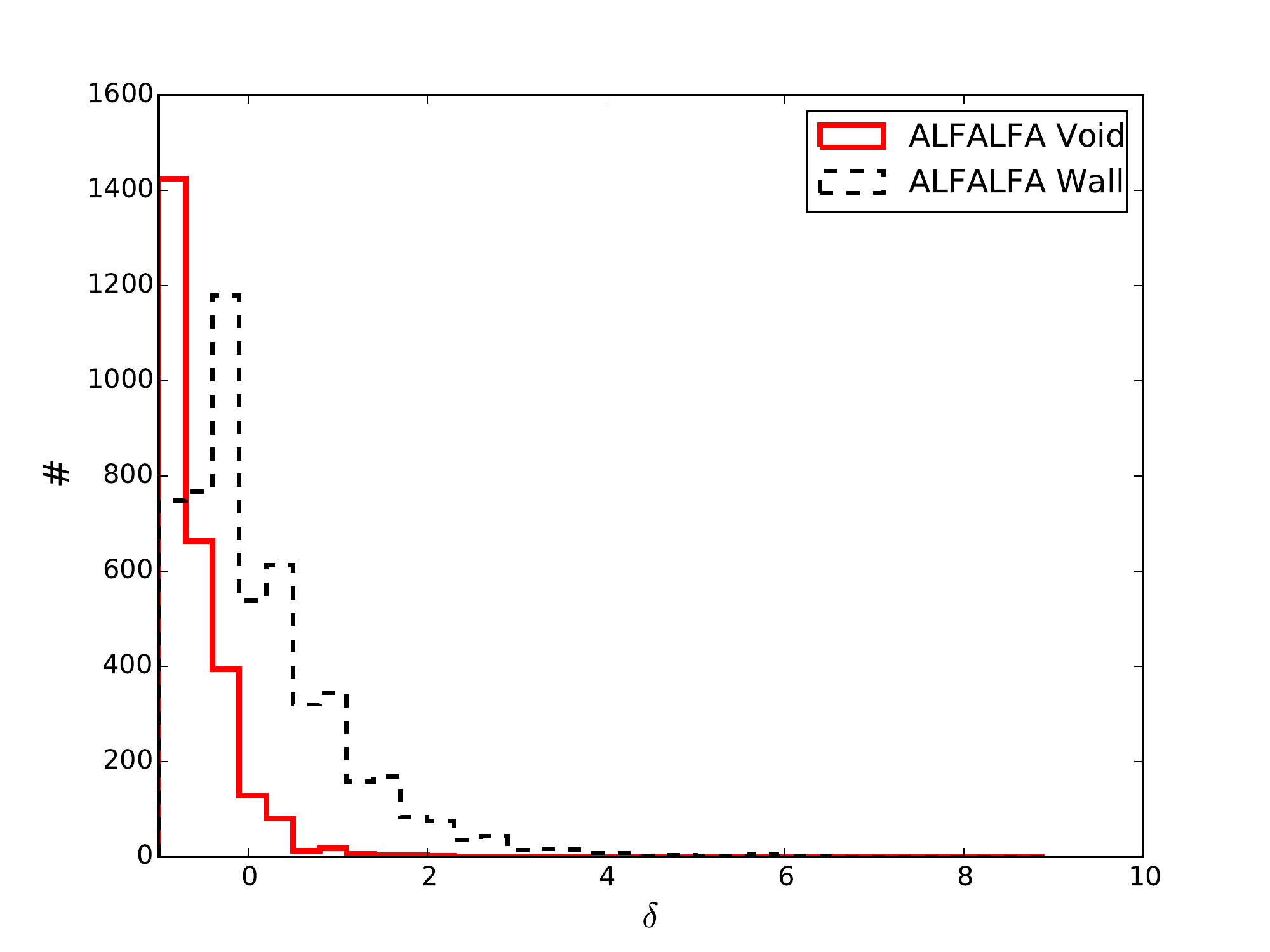} 
  \end{tabular}
\caption[Local densities of NSA and ALFALFA galaxies]{
Distribution of relative local densities for void and wall galaxies in the NSA (left) and ALFALFA (right) samples. 
Void and wall galaxies were selected using a volume-limited sample based on $M_r=-20.1$, 
whereas the small scale densities were calculated using a volume-limited sample based on $M_r=-18.5$. 
We see that wall galaxies range anywhere from isolated to clustered regions. 
Our void galaxies are grouped together with relatively fainter galaxies ($-20.1<M_r<-18.5$) within the voids on smaller scales than are probed by the void catalog.}
\label{fig:density_hist}
\end{figure}
Previous work indicates that star formation properties may be influenced by local density. 
For example, \cite{Elbaz2007} find that galaxy SSFRs decrease with stellar mass and increase with density at $z\sim1$. 
\cite{Huang2012Dwarf} find that low \hi\ mass ALFALFA detections ($M_{HI}<10^{7.7}$) 
in the Virgo Cluster have gas depletion time scales less than a Hubble time. This implies that 
either the dwarf galaxies are undergoing gas-stripping interactions (common to galaxy clusters) or 
the highest density regions are somehow enhancing SFEs. 

In Section \ref{sec:resultsII}, we study the effects of local density on SFE in ALFALFA galaxies. 
We estimate the small-scale densities of each galaxy within the volume covered by VoidFinder, 
based on a volume-limited sample from the SDSS DR7 KIAS-VAGC \citep{Choi2010} catalog. 
We create a volume-limited sample with an absolute magnitude cut 
of $M_r=-18.5$ corresponding to a redshift cut of $z=0.06$ 
to ensure that we completely enclose the NSA volume ($z_{max}=0.055$). 
We apply a nearest-neighbors algorithm to the volume-limited sample with a smoothing scale of $\sim2.5$ h$^{-1}$Mpc 
to approximate the local density, $\rho$, within the volume covered. 
We then evaluate the relative local density  
of each galaxy in the NSA catalog mentioned above.  
Figure \ref{fig:density_hist} shows the distribution of our void and wall galaxies as a function of 
small-scale density contrast, $\delta=(\rho-\bar{\rho})/\bar{\rho}$. 
We find that void galaxies typically lie in regions where the local density contrast is less than the mean, 
but there may be filaments or groups of $M_r\le-18.5$ galaxies within the voids 
increasing the local densities within the voids. 
Additionally, we find wall galaxies with small $\delta$ that are 
not located in large-scale voids.

\section{ESTIMATING STAR FORMATION PROPERTIES}
\label{sec:methII}
To determine if large-scale environment affects star formation in galaxies, and on what time scales, 
we measure the star formation rate (SFR) of galaxies using two independent methods.  
The first method, described in detail in \cite{Salim2007}, uses $FUV$ photometry. 
$FUV$ photometry depends primarily on O- and B-type stars with masses $M\gtrsim 3M_{\sun}$ 
and lifetimes less than $\sim 300$ Myr. 
Therefore, the $FUV$ method best estimates 
the star formation rates of galaxies over the past $\sim100$ Myr. 
The second method is described in \cite{Lee2009} and measures SFRs using \Ha\ spectral lines. 
Only massive ($>10$ $M_{\sun}$), young ($<20$ Myr) stars (O-type stars) significantly contribute to the 
integrated ionizing flux of a galaxy; 
thus, the \Ha\ method provides a good estimate of SFRs on short timescales ($\sim10$ Myr). 
Both of these methods use the star formation relations of \cite{Kennicutt1998}. 
We briefly describe the methods in this section.

\subsection{$FUV$ Method}
\label{sec:sfr_fuv}
We measure the SFRs of galaxies over the last 100 Myr using the 
GALEX photometric $FUV$ fluxes from the NSA catalog. 
$FUV$ photometry is sensitive to dust; therefore, we 
correct the rest frame $FUV$ fluxes for dust attenuation. 
We do so via the empirical equations found from SED fitting of GALEX galaxies in \cite{Salim2007}. 
These authors find the effects of dust are stronger for red ($NUV-r\ge4$) galaxies than blue ($NUV-r<4$) galaxies. 
Thus red galaxies are corrected by 
\begin{equation}
A_{FUV}= 
\begin{cases}
3.32(FUV-NUV)+0.22 & FUV-NUV \le 0.95 \\ 
3.37 & FUV-NUV > 0.95 
\end{cases},
\label{eq:red_fuv_dust}
\end{equation}
and blue galaxies are corrected by 
\begin{equation}
A_{FUV}= 
\begin{cases}
2.99(FUV-NUV)+0.27 & F-N \le 0.90\\ 
2.96 & F-N > 0.90
\end{cases}.
\label{eq:blue_fuv_dust}
\end{equation} 
Here, $A_{FUV}$ is the dust attenuation correction for SFRs obtained using the $FUV$ method. 
For galaxies with dust attenuation corrections falling below zero, we set the correction to $A_{FUV}=0$, 
to ensure that we are not 
artificially adding dust back into the system. 
We apply this correction to the rest-frame flux, $f^0$, in the following way: 
\begin{equation}
f_{FUV}=f^{0}10^{A_{FUV}/2.5}.
\label{eq:f_fuv}
\end{equation}

From the dust corrected $FUV$ fluxes, $f_{FUV}$, we calculate the $FUV$ luminosities via 
\begin{equation}
L_{FUV}= 4\pi D^{2}_{L}f_{FUV}, 
\label{eq:L_fuv}
\end{equation}
where $D_L$ is the luminosity distance. 
Following \cite{Salim2007}, we then apply the \cite{Kennicutt1998} SFR relation with a factor 
to better match the stellar evolution models of \cite{Bruzual2003}, 
\begin{equation}
SFR= 1.08\times 10^{-28}L_{FUV}, 
\label{eq:fuv_sfr}
\end{equation}
to obtain the average SFR in units of  $M_{\sun}$yr$^{-1}$ over the past $\sim100$ Myr.

\subsection{\Ha\ Method}
\label{subsec:Ha_sfr}
For an estimate of galaxy SFRs 
over the last $\sim10$ Myr, 
we calculate the \Ha\ luminosity and 
apply the \cite{Kennicutt1998} SFR relation similar to equation (\ref{eq:fuv_sfr}): 
\begin{equation}
SFR= 7.9\times 10^{-41.28}L_{{\text\Ha}} 
\label{eq:ha_sfr}
\end{equation}
to obtain the average SFR in units of  $M_{\sun}$yr$^{-1}$ over the past $\sim10$ Myr. 
Here, $L_{{\text\Ha}}$ is the \Ha\ luminosity obtained from 
\begin{equation}
L_{\text\Ha}= 4\pi D^{2}_{L}f_{\text\Ha}, 
\label{eq:L_Ha}
\end{equation}
where $f_{\text\Ha}$ is the dust-corrected \Ha\ flux and $D_L$ is the luminosity distance.  
As with the $FUV$ method above, we must make dust attenuation corrections to 
the \Ha\ SFR estimates. To do so, we use the correction suggested in \cite{Lee2009}: 
\begin{equation} 
A_{\text \Ha}= 5.91\log(\text{\Ha/H$\beta$})-2.70. 
\label{eq:ha_dust}
\end{equation} 

Additionally, we make an aperture correction to the \Ha\ luminosity to adjust for the 
3'' diameter SDSS fiber potentially being smaller than the size of the galaxy observed. 
The SDSS fiber detects \Ha\ emission from a fraction 
of the galaxy that depends on the redshift and intrinsic size of the galaxy.  
We use the correction of \cite{Hopkins2003} adjusting the 
\Ha\ luminosity by a factor of $10^{-0.4 (r_\text{Petro}-r_\text{fiber})}$. 
Here, $r_{Petro}$ is the $r$-band Petrosian magnitude of the full galaxy, and 
$r_\text{fiber}$ is the $r$-band magnitude within the fiber. 
This aperture correction assumes that star formation is uniformly distributed across the galaxy. 
Depending on galaxy orientation and non-uniform distribution of star formation activity, 
\cite{Hopkins2003} find that this aperture correction may over estimate the SFR 
by up to 15\% for galaxies out to $z\sim0.2$.

\subsection{Star Formation Rates} 
\label{subsec:meth_sfr} 
\begin{figure*}[ht]
\centering
  \includegraphics[scale=0.5,trim = 0cm 0.5cm 0.cm 1.25cm, clip=True]{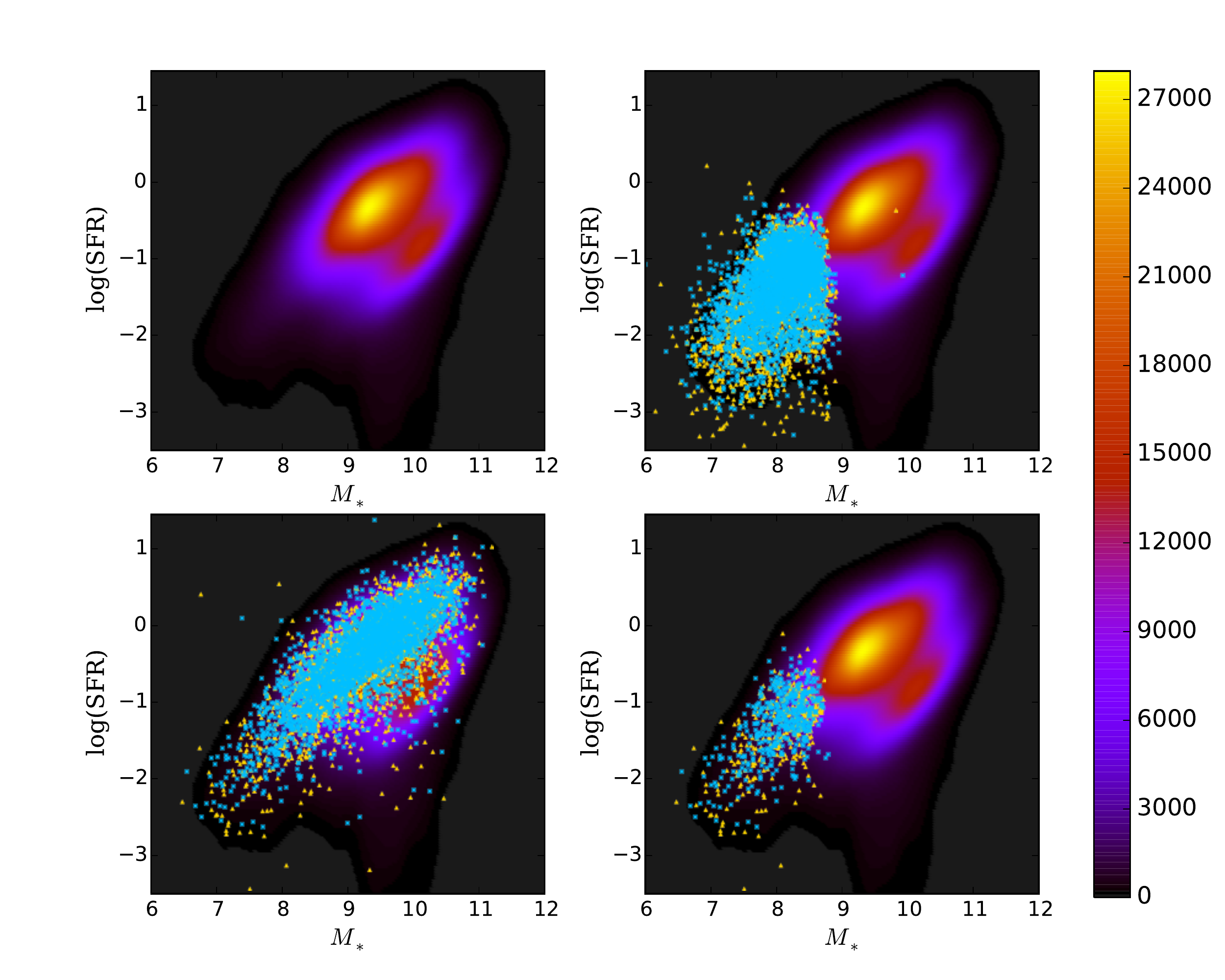} 
  \caption[SFR vs. $M_r$]{ 
  Upper Left: Color contours depict the stellar mass, $\log(M_*)$, vs log(SFR$_{FUV}$) distribution for all NSA galaxies. 
  The colorbar indicates the number of galaxies in the sample in this space. 
  Upper Right: NSA Dwarf void (blue crosses) and 
  stellar mass-matched wall (gold triangles) galaxies overplotted on the full NSA distribution. 
  SFR is closely correlated with mass; therefore, it is unsurprising that dwarf galaxies tend to have low SFRs. 
  Most dwarf galaxies in the NSA are star forming with a small fraction appearing in an extension of the passive sequence. 
  Lower Left: All ALFALFA void and stellar mass-matched wall galaxies. 
  ALFALFA typically detects active star forming galaxies, mostly avoiding the passive galaxy region in the plot. 
  Lower Right: ALFALFA Dwarf void and stellar mass-matched wall galaxies. 
  These galaxies lie almost exclusively in the star forming sequence. }
 \label{fig:sm_sfr}
\end{figure*}
We calculate the SFR for both methods mentioned above for all galaxies in our data set. 
The NSA catalog includes estimates of \Ha\ line flux and $NUV$/$FUV$ photometry for all galaxies. 
Thus, 
\FloatBarrier \noindent
we did not exclude low SFR galaxies and there is no bias in the sample toward high SFR galaxies. 
In this work, we investigate the differences of the star formation properties based on environment 
and sample selection of the following samples: 
all galaxies identifiable as a void or wall galaxy (hereafter, NSA), 
all dwarf galaxies identifiable as a void or wall galaxy (NSA Dwarf), 
galaxies within the aforementioned NSA sample with ALFALFA \hi\ masses measured with S/N$>$4.5 (ALFALFA), 
and galaxies within the NSA Dwarf sample with corresponding \hi\ clouds with S/N$>$4.5 (ALFALFA Dwarf). 

We investigate the effects of sample selection on SFR by comparing the location of galaxy subsamples within 
log(SFR$_{FUV}$) vs. stellar mass (log$M_*$) space to the full galaxy catalog in Figure \ref{fig:sm_sfr}. 
Again, the NSA sample is a superset of all other galaxy samples used in this work (i.e. 
the NSA Dwarf, ALFALFA, and ALFALFA Dwarf samples). 
The upper left panel of Figure \ref{fig:sm_sfr} depicts a density contour of the NSA sample in 
a log(SFR$_{FUV}$) vs. log($M_*$) diagram. 
We see a clear bimodality in the  diagram, indicating 
the existence of both an ``active'' (higher SFR) population depicted by the main peak in the contour diagram,  
and a ``20pt'' (lower SFR) population depicted by the smaller peak to the lower right of the main peak. 
In the lower left panel, we select galaxies within the NSA catalog with \hi\ emission lines strong enough to be detected by ALFALFA. 
We divide the ALFALFA detections into void (blue crosses) and wall (yellow triangles) galaxies and overplot them on  
the NSA density contour in the log(SFR) vs. log($M_*$) space. 
This panel shows the overall effects of \hi\ selection on galaxy SFR.
We notice that we lose most evidence of bimodality in the distribution. 
This confirms that \hi\ surveys tend to detect galaxies that are actively forming stars today. 
These findings corroborate the work of \cite{Huang2012}, who mention that the ALFALFA galaxies have 
higher SFRs than optically selected samples from the SDSS.   
We see some evidence at the high stellar mass end that a population of quiescent galaxies exists 
that contain just enough gas to be detected by ALFALFA. 
Using the \Ha\ method rather than the $FUV$ method, 
we find very similar results on the sample selection effects and dwarf selection effects; 
thus, we will not plot the \Ha\ estimated SFR distributions here.

\section{SPECIFIC STAR FORMATION RATES}
\label{sec:resultsI}
\begin{figure*}[ht]
    \centering
      \includegraphics[scale=0.5,trim = 0cm 0.5cm 0.cm 1.25cm, clip=True]{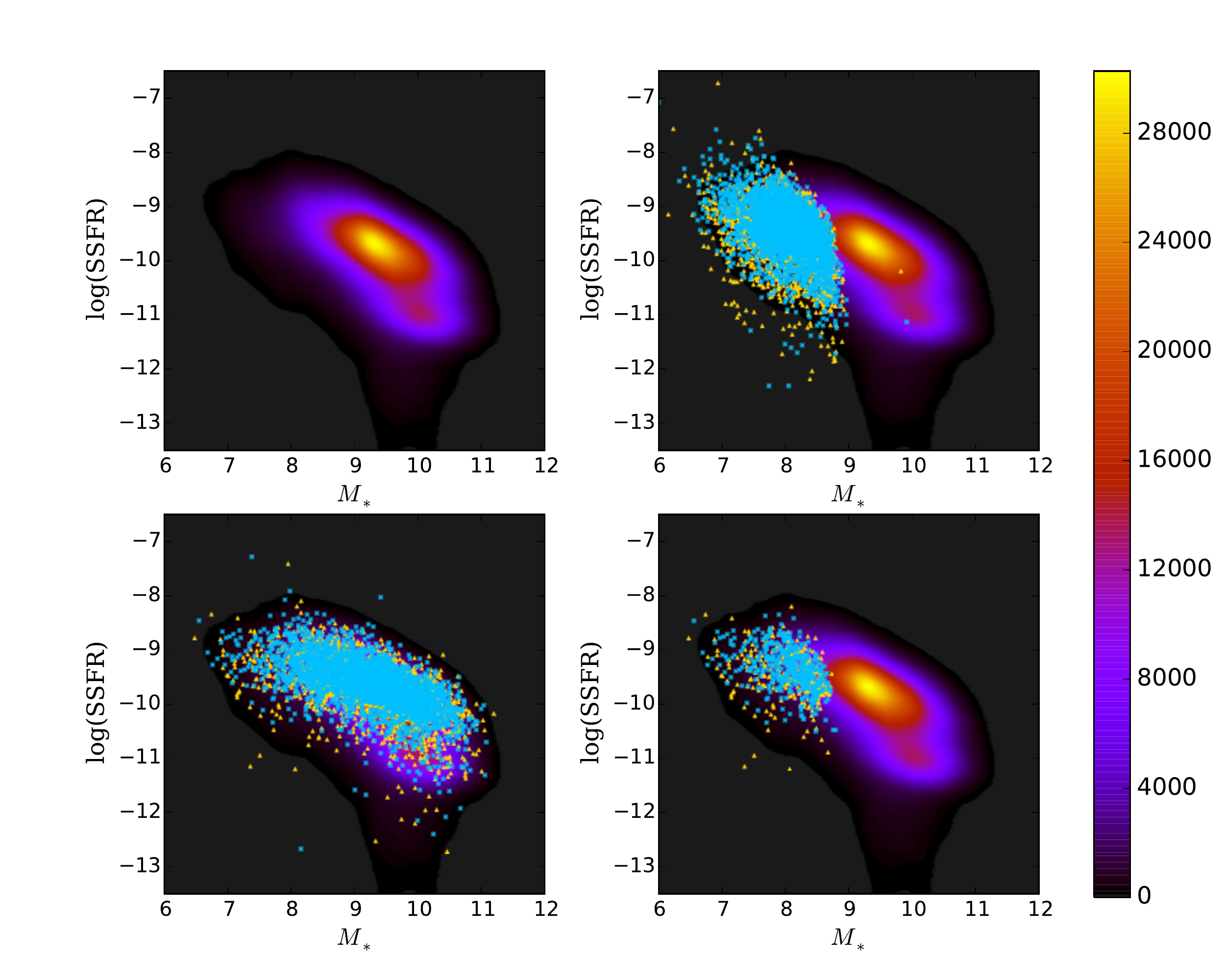} 
	\caption[SSFR vs. $M_r$]{ 
	Upper Left: Color contours depict the stellar mass, $\log(M_*)$, vs log(SSFR$_{FUV}$) distribution for all NSA galaxies. 
	The colorbar indicates the number of galaxies in the sample in this space. 
	Upper Right: NSA Dwarf void (blue crosses) and wall (gold triangles) galaxies overplotted on the full NSA distribution. 
	The wall dwarf galaxies have been downsampled so that the wall dwarf galaxy stellar mass distribution matches 
	the stellar mass distribution of void dwarf galaxies.  
	Dwarf galaxies tend to have relatively high SFRs given their size. 
	A majority of dwarf galaxies lie in the active star forming region in this space. 
	Lower Left: All ALFALFA void and stellar mass-matched wall galaxies. As seen with the SFR distributions, 
	ALFALFA typically detects active star forming galaxies, avoiding the passive galaxy region in the plot. 
	Lower Right: ALFALFA void dwarf and stellar mass-matched wall dwarf galaxies. 
	These galaxies are actively forming stars at relatively high rates. }
     \label{fig:nsa_ssfr_contours}
\end{figure*}
SFR is correlated with stellar mass; therefore it is no surprise that  
selecting only the dwarf galaxies (right column of Figure~\ref{fig:sm_sfr})
produces lower SFRs than the average galaxies in the full catalogs. 
To account for the correlation between the SFR and mass of a galaxy, 
we normalize the SFR of each object by its stellar mass 
to obtain its specific SFR (SSFR=SFR/$M_*$), which is a measure of the SFR per unit mass. 
Figure \ref{fig:nsa_ssfr_contours} depicts the location of each galaxy sample within the SSFR$_{FUV}$ vs. stellar mass 
plane. The upper left panel shows the density contour of the NSA sample. 
The passive galaxy sequence is seen primarily as an extension from the main active sequence moving towards 
lower SSFRs around $\log(M_*)\sim10$ and continuing towards the bottom edge of the figure. 
This density contour is replotted in the background of each subplot of the figure 
for the sake of comparing how sample selection affects the SSFRs. 
In the lower left panel,  
we divide the ALFALFA detections into void (blue crosses) and wall (yellow triangles) galaxies and scatter them over  
the NSA density contour in the SSFR vs. stellar mass space. 
This panel shows the overall effects of \hi\ selection on galaxy SSFR. 
As seen in Section \ref{subsec:meth_sfr}, we find the ALFALFA sample lies 
primarily in the star forming sequence and is sparse in the 
passive galaxy region. \cite{Huang2012} find a similar result for all ALFALFA galaxies. 

In the right column of Figure~\ref{fig:nsa_ssfr_contours}, we select out only the dwarf galaxies ($M_r\ge-17$) within the 
NSA (top right) catalog and overplot them on the full NSA density contours. 
We find that, as a whole, dwarf galaxies tend to lie primarily in the active star forming sequence, 
with a few trailing into the passive region. Dwarf galaxies have higher SSFRs than average galaxies as 
predicted by the high resolution hydrodynamic simulations of \cite{Cen2011}.    
We then select out the dwarf galaxies that have an \hi\ flux large enough to be detected in ALFALFA. 
We plot these galaxies in the bottom right panel of Figure \ref{fig:nsa_ssfr_contours}, 
and find that the \hi\ selection cuts out almost all of the dwarf galaxies lying in the passive galaxy region.  

\begin{figure*}[ht]
	\includegraphics[scale=0.41,trim = 0cm 0cm 4cm 1cm, clip=True]{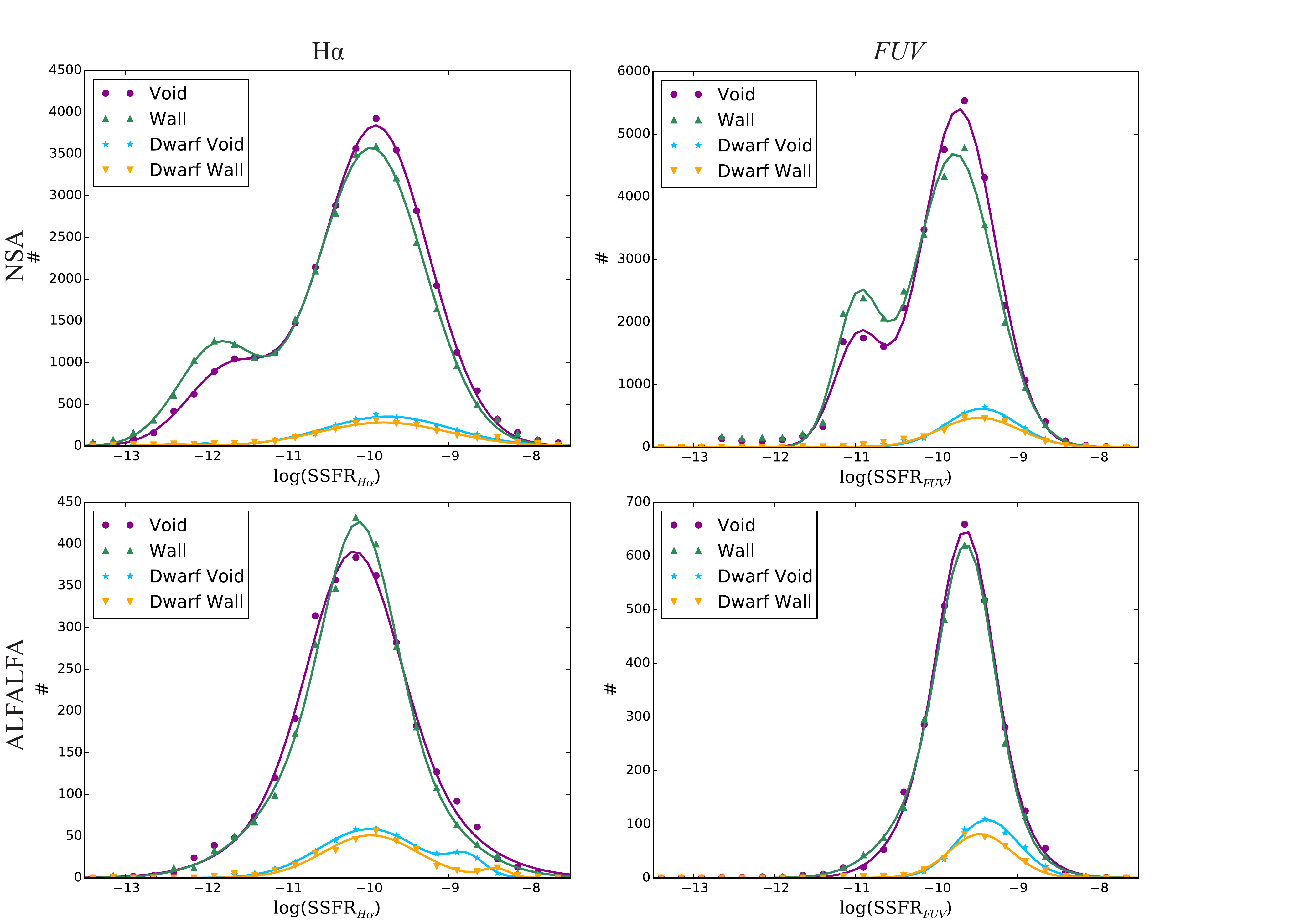}
	\caption[\Ha\ and $FUV$ SSFR fits]{ 
	SSFR distribution of void and stellar mass-matched wall galaxies measured via \Ha\ and $FUV$ methods for 
	NSA, NSA Dwarf, ALFALFA, and ALFALFA Dwarf samples. Lines denote the best-fit summed Gaussian. 
	Upper Left: NSA SSFR distributions via the \Ha\ method. Void galaxies have higher SSFRs than wall galaxies 
	in the full sample as well as the dwarf sample.  
	Upper Right: NSA SSFR distributions via the $FUV$ method. 
	Again, the void galaxies have higher SSFRs than wall galaxies in the full sample as well as the dwarf sample. 
	Lower Left:  ALFALFA SSFR distributions via the \Ha\ method. 
	Due to \hi\ selection effects, we do not see evidence of the passive galaxy sequence. 
	We notice no difference in the full void and wall distributions. 
	A peak in the high-SSFR end of the void dwarf distribution reveals a population of 
	star-bursting void dwarf galaxies. 
	Lower Right:  ALFALFA SSFR distributions via the $FUV$ method. 
	There is no significant difference between the full void and wall samples. 
	The void dwarf distribution appears to be shifted towards higher SSFRs.}  
     \label{fig:alfnsa_ssfr_hists}
\end{figure*} 
To quantitatively determine the large-scale environmental effects on the SSFR,
we fit the $FUV$ and \Ha\ SSFRs for the NSA and ALFALFA samples by two summed Gaussians: 
\begin{equation}
f(x) = \frac{A_1}{\sigma_1\sqrt{2\pi}} e^{-\frac{(x-\mu_1)^2}{2\sigma_1^2}}+
			\frac{A_2}{\sigma_2\sqrt{2\pi}} e^{-\frac{(x-\mu_2)^2}{2\sigma_2^2}}.
\label{eq:double_gaussian}
\end{equation}
Here, $A$ is the amplitude, $\mu$ is the mean, and $\sigma$ is the standard deviation of each 1-D gaussian. 
The two Gaussians used to fit the SSFR distributions correspond to 
the passive and active star forming populations of each sample. 
Figure \ref{fig:alfnsa_ssfr_hists} depicts the SSFR distributions with best-fitting summed Gaussians. 
See Table \ref{tab:ssfr_fits} for the best fitting Gaussian parameters for each sample. 
\begin{deluxetable}{lrlcccccc} 
	\tablewidth{0pt}
	\tablecaption{Gaussian fits to the SSFRs\label{tab:ssfr_fits}}
	\tablehead{\colhead{Sample}&\colhead{type}&\colhead{method}&\colhead{$A_1$}&\colhead{$\mu_1$}&\colhead{$\sigma_1$}&\colhead{$A_2$}&\colhead{$\mu_2$}&\colhead{$\sigma_2$}}
	\startdata
	NSA				& void	&	\Ha		&	3841&	-9.906&	0.651&	926&	-11.707&	0.518\\ 
	NSA				& wall	&	\Ha		&	3576&	-9.961&	0.658&	1194&	-11.858&	0.489\\ 
	NSA	Dwarf		& void	&	\Ha		&	355&	-9.752&	0.773&	33&		-12.029&	0.078\\ 
	NSA	Dwarf		& wall	&	\Ha		&	282&	-9.803&	0.793&	21&		-12.401&	0.261\\ 
	NSA				& void	&	$FUV$	&	5406&	-9.719&	0.454& 	1727&	-10.963&	0.294\\ 
	NSA				& wall	&	$FUV$	&	4691&	-9.769&	0.489&	2269&	-10.970&	0.272\\ 
	NSA Dwarf		& void	&	$FUV$	&	612&	-9.434&	0.438&	---&		---&		---\\ 
	NSA	Dwarf		& wall	&	$FUV$	&	471&	-9.467&	0.488&	---&		---&		---\\ 
	ALFALFA 			& void	&	\Ha		&	263&	-10.177&	0.540&	129&	-10.110&	0.991\\ 
	ALFALFA 			& wall	&	\Ha		&	167&	-10.199&	0.896&	260&	-10.104&	0.429\\ 
	ALFALFA Dwarf	& void	&	\Ha		&	59&		-9.984&	0.618&	22&		-8.781&	0.215\\ 
	ALFALFA Dwarf	& wall	&	\Ha		&	51&		-9.968&	0.579&	11&		-8.423&	0.163\\ 
	ALFALFA 			& void	&	$FUV$	&	416&	-9.633&	0.323&	233&	-9.697&	0.561\\ 
	ALFALFA 			& wall	&	$FUV$	&	449&	-9.616&	0.331&	181&	-9.814&	0.618\\ 
	ALFALFA Dwarf	& void	&	$FUV$	&	104&	-9.378&	0.376&	     5&	-9.114&	0.570\\ 
	ALFALFA Dwarf	& wall	&	$FUV$	&	81&		-9.456&	0.383&	     1&	-8.218&		3.131\\ 
	\enddata
\end{deluxetable}

The left panels in Figure~\ref{fig:alfnsa_ssfr_hists} plot SSFR distributions 
averaged over the last $\sim10$ Myr, estimated using the \Ha\ line. 
The right panels plot the SSFR distributions averaged over the last $\sim100$ Myr, 
estimated using the $FUV$ photometry. 
The top row shows the stellar mass-matched NSA data, and the bottom row shows the stellar mass-matched ALFALFA data. 
It is obvious from the top left panel of Figure \ref{fig:alfnsa_ssfr_hists} 
that void galaxies have higher SSFRs than wall galaxies in both the 
active (13\% shift) and passive (41\% shift) regions based on the \Ha\ estimates. 
For the NSA Dwarf sample, the void galaxy distribution has an 11\% shift towards higher SSFRs 
than stellar mass-matched wall dwarf galaxies when using the \Ha\ method. 

In the upper right panel of Figure~\ref{fig:alfnsa_ssfr_hists}, we measure the $FUV$ SSFR for NSA galaxies and find that 
void galaxies have higher SSFRs than stellar mass-matched wall galaxies in both 
the active (13\%) and passive (41\%) populations. 
The NSA Dwarf void galaxies also experience higher SSFRs based on the $FUV$ method by $\sim7\%$. 
Based on the $FUV$ method, we find that 90\% of void dwarf galaxies and 85\% of wall dwarf galaxies have 
SSFRs high enough to double their stellar mass in a Hubble time (corresponding to log(SSFR)$>-10.1$ $M_{\sun}$yr$^{-1}$).  
Computing the Kolmogorov-Smirnov (K-S) test statistic on the mass-matched void and wall SSFR 
distributions on both time scales returns a $p$-value less than 0.001 for the NSA void and stellar mass-matched wall samples 
as well as for the NSA Dwarf void and stellar mass-matched wall samples. 
Thus, we can reject the null \FloatBarrier\noindent
hypothesis that the void and stellar mass-matched wall galaxies from 
these two sets were drawn from the same distribution. 

\subsection{Tracking the Sample Selection Effects}
\label{subsec:alf_ssfr}
SSFRs are highly anti-correlated with $NUV-r$ color 
(with correlation coefficients of r$_s$=-0.923 for the NSA sample and r$_s$=-0.875 for the ALFALFA sample). 
That is, bluer galaxies tend to have higher SSFRs. 
Of course, there are always exceptions, such as the population of dust-reddened 
spiral galaxies detected by ALFALFA (e.g. \cite{Moorman2015}). 
Edge-on, blue spiral galaxies will have high SSFRs, but will generally appear redder than they truly are 
due to dust in the galaxy absorbing and scattering photons with shorter wavelengths.  
Because of the bias towards blue galaxies in the \hi\ samples, 
we will see hardly any difference between the SSFRs of void and wall galaxies. 
On average, the SSFRs of an \hi\ selected sample should be 
higher than for an optically selected sample because we are lacking galaxies from the red, passive galaxy sequence. 
While the average SSFRs are higher, comparing only the active star-forming sequence of ALFALFA to that of NSA 
reveals a shift towards lower SSFRs in \hi\ surveys. 
This is unsurprising given that the ALFALFA blue cloud is shifted 
towards redder colors than the SDSS blue cloud (see, for instance, fig. 1 in \cite{Moorman2015}). 
The shift is possibly indicative of \hi\ selected galaxies being more  
dust-reddened on average than optically selected star forming galaxies. 

In the bottom panels of Figure \ref{fig:alfnsa_ssfr_hists}, there are no
discernible differences between the full ALFALFA void and ALFALFA stellar mass-matched wall SSFR 
distributions on either time scale. 
Estimating a shift in the void and wall distributions based on the summed Gaussian fits, 
we find only a 5\% shift towards lower \Ha\ SSFRs with a broader distribution in voids 
than walls (see Table \ref{tab:ssfr_fits} for fits). 
In the bottom right panel, we find a $<1\%$ shift towards higher $FUV$ SSFRs and no width differences in the distributions, 
implying that the SSFRs of \hi\ selected galaxies are not dependent on large scale structure based on the $FUV$ method.   
The K-S test results in $p_{K-S}=0.09$ for \Ha\ SSFRs and $p_{K-S}=0.14$ for $FUV$ SSFRs, 
meaning that we cannot reject that the ALFALFA void and wall samples were drawn from the 
same distribution on either time scale.  

Limiting our scope to the ALFALFA Dwarf sample, we find in the \Ha\ SSFR distribution 
that the main peak of the void sample, centered on log(SSFR)$\sim-10$, 
is shifted only by $\sim4\%$ towards lower SSFRs than the main peak of the wall galaxy distribution. 
Both the void and wall ALFALFA Dwarf samples have secondary peaks at the 
high end of the \Ha\ SSFR distribution not apparent in any other sample. 
This is indicative of a population of star bursting dwarf galaxies that has been 
active very recently (10 Myr) especially in voids. 
The wall galaxies, however, seem to be lacking a sample of high \Ha\ SSFRs 
around $-9.2<$log(SSFR)$<-8.4$ where the secondary void peak is strongest. 
This effect is most likely due to ALFALFA's bias towards detecting galaxies preferentially found in voids.  
At higher SSFRs, the wall dwarf 
distribution increases again, to closely match the void distribution around $\log($SSFR$)\sim-8.5$. 
Due to these high-SSFR features, a K-S test yields $p_{K-S}=0.005$, 
indicating that we should reject that these samples are drawn from the same population.

On longer star formation time scales ($\sim100$ Myr), the ALFALFA Dwarf void galaxies are shifted 
towards higher $FUV$ SSFRs by 23\% with $p_{K-S}=0.002$. 
Based on this test statistic, we should reject that the void and wall ALFALFA Dwarf detections 
were drawn from the same distribution. 
Based on these SSFR results, we expect that we will find the 
star formation efficiency of ALFALFA void and wall galaxies to be very similar for both the \Ha\ and $FUV$ methods.
We also suspect the ALFALFA Dwarf galaxies will vary only slightly on both short and long time scales.  
Compared to the NSA Dwarf SSFR$_{FUV}$ distribution, the ALFALFA Dwarf galaxies 
have higher SSFRs on average with 94\% of void dwarfs and 93\% of wall dwarfs having 
SSFRs high enough to double their stellar mass within a Hubble time \citep{Cen2011}.

\section{STAR FORMATION EFFICIENCY}
\label{sec:resultsII}
\begin{figure}[ht]
	\includegraphics[scale=0.4235,trim = 0cm 0cm 3cm 0cm, clip=True]{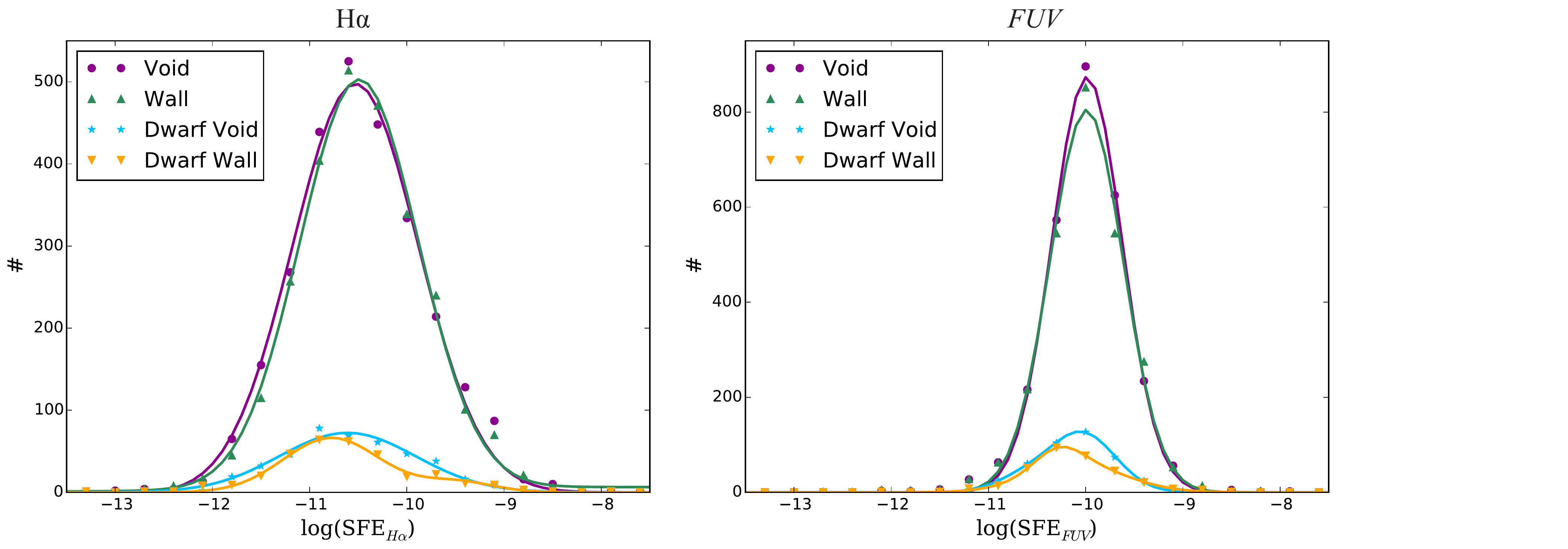}
	\caption[\Ha\ and $FUV$ SFE fits]{
	Left: Distribution of the \Ha\ SFE of all void and wall galaxies as well as the dwarf void and wall galaxies. 
	Right: Distribution of the $FUV$ SFE of all void and wall galaxies as well as the dwarf void and wall galaxies. 
	There is no discernible difference between the full void and wall SFE distributions of ALFALFA galaxies on either timescale. 
	Isolating the ALFALFA dwarf galaxies reveals a slight trend towards higher SFEs in void galaxies. }
  \label{fig:alf_sfe_hists}
\end{figure}
\begin{deluxetable}{lrlcccccc} 
	\tablewidth{0pt}
	\tablecaption{Gaussian fits to the SFEs\label{tab:sfe_fits}}
	\tablehead{\colhead{Sample}&\colhead{type}&\colhead{method}&\colhead{$A_1$}&\colhead{$\mu_1$}&\colhead{$\sigma_1$}&\colhead{$A_2$}&\colhead{$\mu_2$}&\colhead{$\sigma_2$}}
	\startdata
	ALFALFA 			& void	&	\Ha		&	500&	-10.530&		0.646&	-3&		-11.558&	1.138\\ 
	ALFALFA 			& wall	&	\Ha		&	499&	-10.492&		0.607&	7&		-7.618&	1.894\\ 
	ALFALFA Dwarf	& void	&	\Ha		&	72&		-10.611&		0.701&	-2&		-5.956&	0.521\\ 
	ALFALFA Dwarf	& wall	&	\Ha		&	66&		-10.776&		0.493&	13&		-9.488&	0.374\\ 
	ALFALFA 			& void	&	$FUV$	&	874&	-9.986&		0.361&	---&			---&	---\\ 
	ALFALFA 			& wall	&	$FUV$	&	805&	-9.989&		0.3759&	---&			---&	---\\ 
	ALFALFA Dwarf	& void	&	$FUV$	&	125&	-10.038&		0.339&	28&		-10.658&	0.284\\ 
	ALFALFA Dwarf	& wall	&	$FUV$	&	43&		-10.281&		0.240&	56&		-10.082&	0.498\\ 
	\enddata
\end{deluxetable}
A measure of how efficiently galaxies are transforming their gas into stars is the SFR normalized by the \hi\ mass, 
termed the galaxy's star formation efficiency (SFE = SFR/$M_{HI}$). 
In Figure \ref{fig:alf_sfe_hists} we present the SFE distribution of the full ALFALFA void and wall samples and the 
ALFALFA Dwarf void and wall samples for both the \Ha\ method (left panel) and $FUV$ (right panel).  
The lines plotted represent the best-fit summed Gaussians (see Equation~\ref{eq:double_gaussian}) 
with parameters shown in Table \ref{tab:sfe_fits}. 
In the \Ha\ distribution, we find a small (9\%) shift towards lower SFEs and a broader distribution 
in ALFALFA void galaxies compared to walls. 
As with the ALFALFA Dwarf SSFR$_{\text\Ha}$ distributions, 
the SFE$_{\text\Ha}$ wall distribution is lacking at the high SFE end (log(SFE)$\sim-10$). 
This is primarily due to the same large-scale structure (LSS) and \hi\ selection effects discussed in the previous section. 
We also find similar results between the SFE$_{FUV}$ and SSFR$_{FUV}$, 
in that there is no difference in the ALFALFA void and wall distributions and 
there is a lack of ALFALFA Dwarf wall galaxies at high SSFRs. 
This lack of wall galaxies may be due to the prominence of nearby voids within the ALFALFA $\alpha.40$ volume.  
The K-S test statistics ($p_{full}=0.13$ and $p_{dwarf}=0.17$ for the \Ha\ method and 
$p_{full}=0.24$ and $p_{dwarf}=0.11$ for the $FUV$ method) indicate 
that we cannot reject that the samples were drawn from the same distribution. 

\begin{figure*}[ht!]
  \centering
    \begin{tabular}{@{}ccc@{}}  
        \includegraphics[scale=0.45,trim = 0.cm 0.1cm 1.0cm 1.cm, clip=True]{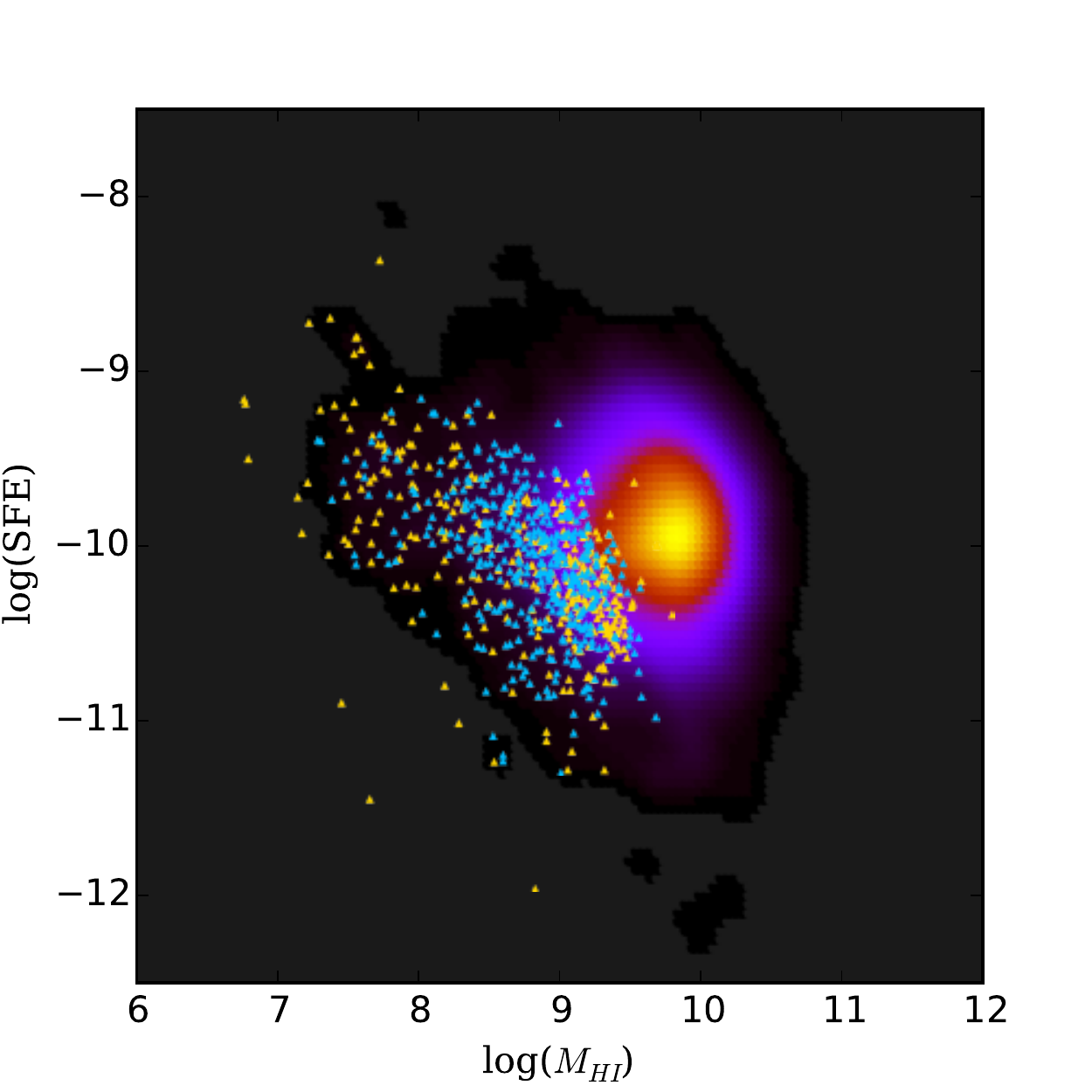} &
        \includegraphics[scale=0.45,trim = 0.cm 0.1cm 1.0cm 1.cm, clip=True]{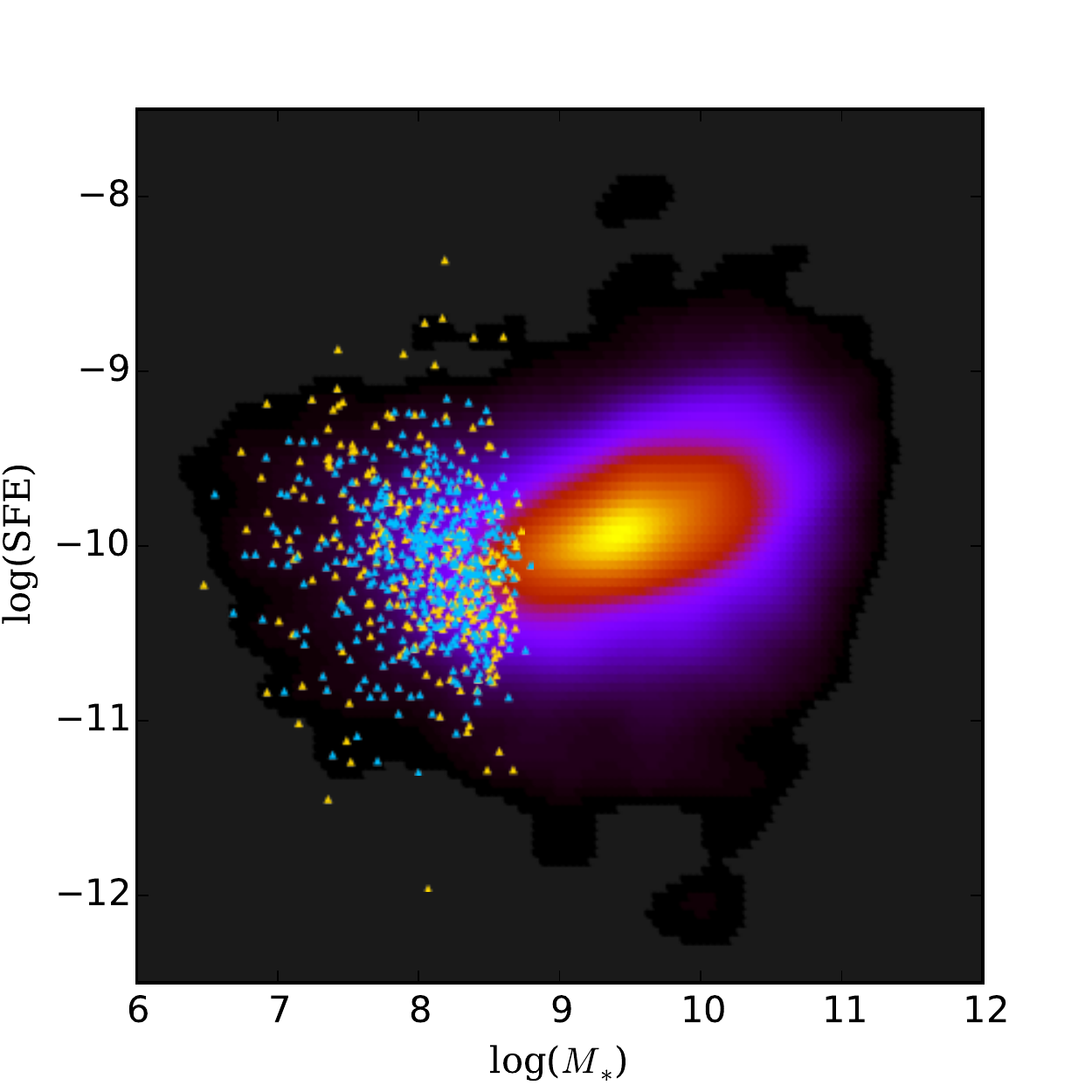} & 
         \includegraphics[scale=0.45,trim = 0.cm 0.1cm 1.0cm 1.cm, clip=True]{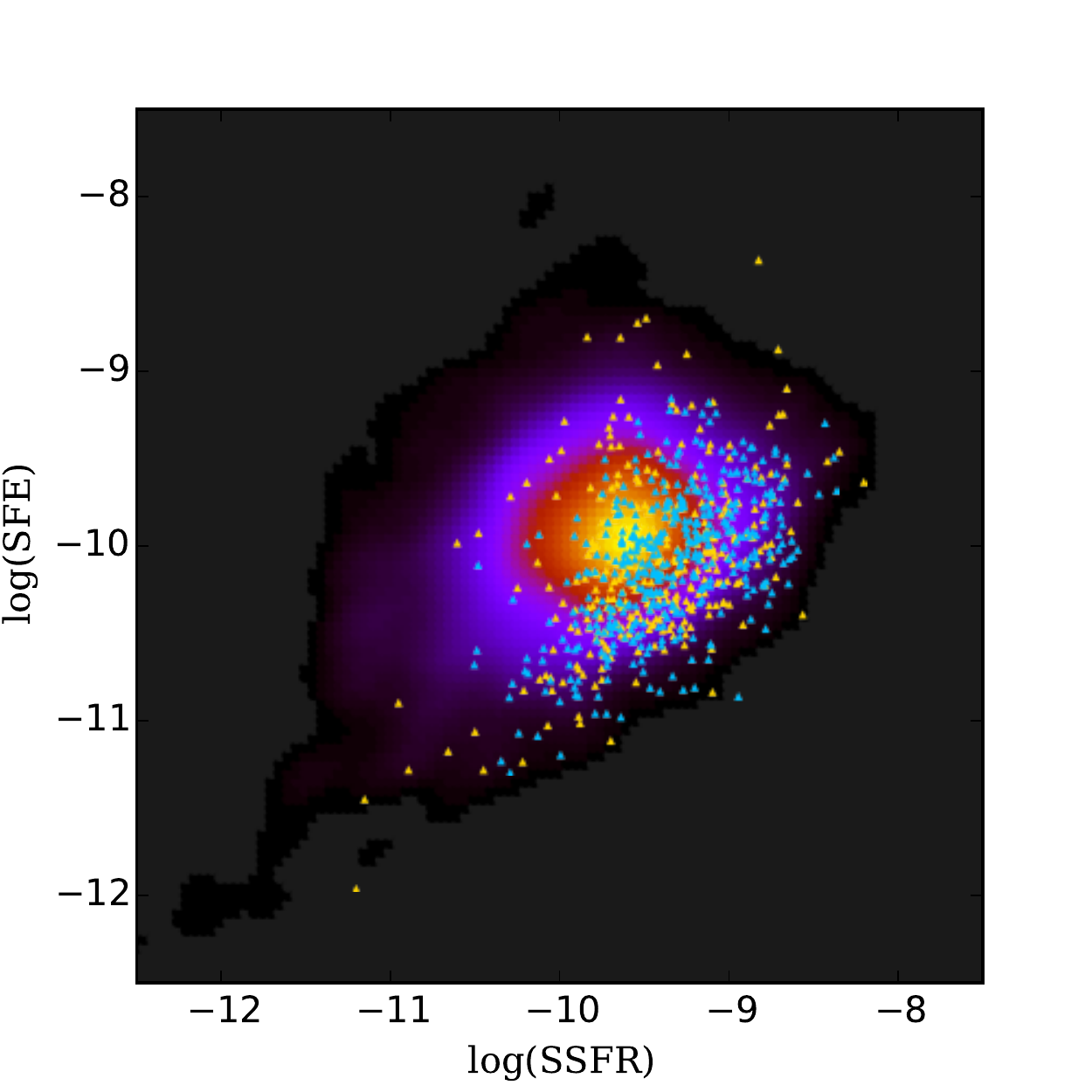}
    \end{tabular}
    \caption[ALFALFA SFE vs. $M_{HI}$,  $M_*$,  and SSFR]{  
    Left to right: ALFALFA galaxy SFEs as a function of $M_{HI}$, $M_*$, and  SSFR. 
    Color contours represent the full ALFALFA sample, while the ALFALFA void dwarf (blue crosses) 
    and ALFALFA wall dwarf (gold triangles) populations are overplotted.
    The wall dwarf galaxies were sampled so that the wall dwarf stellar mass distribution matched the 
    void dwarf stellar mass distribution. 
    We notice a population of wall dwarf galaxies with very high SFEs. } 
  \label{fig:random_alfalfa_images}
\end{figure*}
\begin{figure}[ht]
    \centering
     \includegraphics[scale=0.5,trim = 0.cm 0.1cm 1.0cm 1.cm, clip=True]{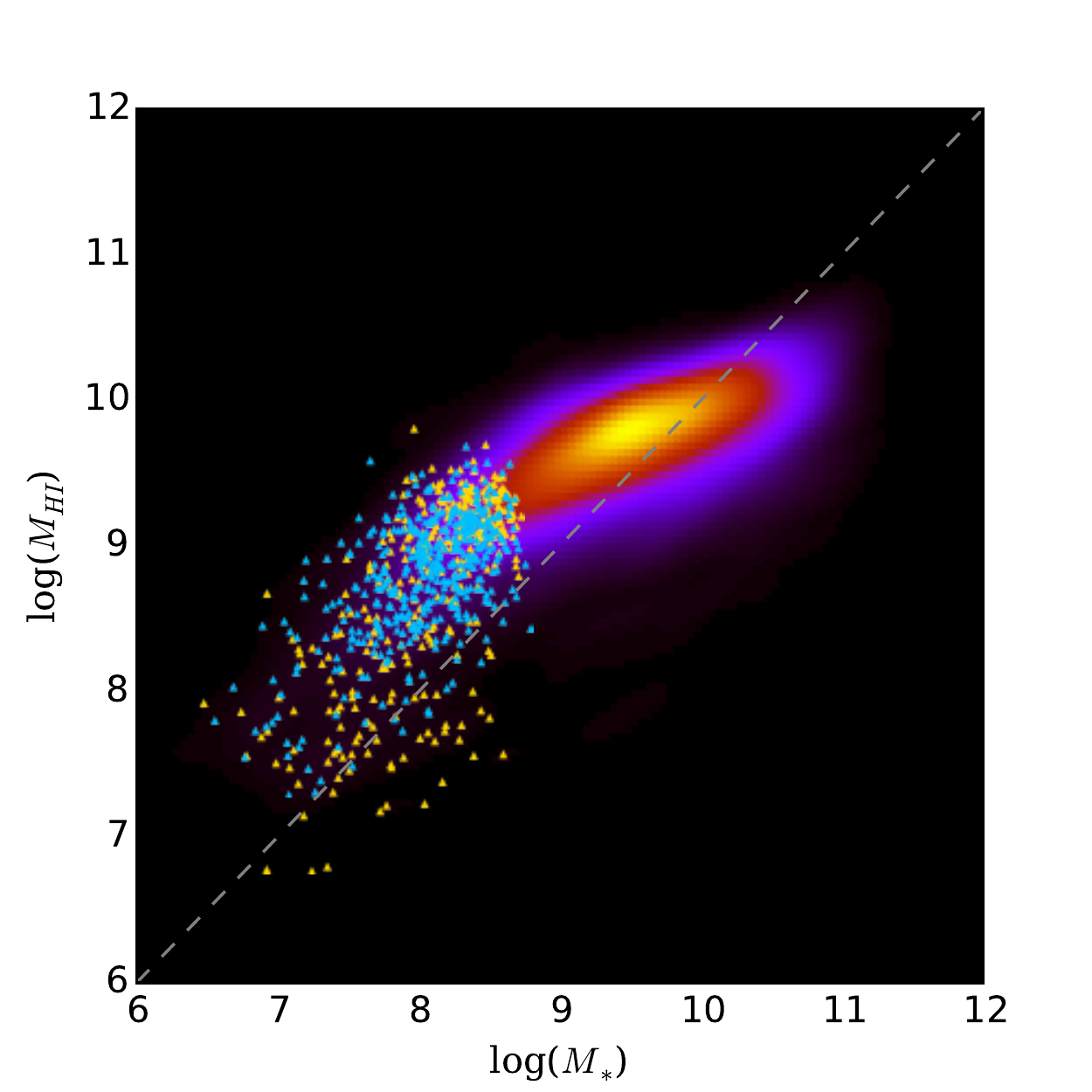} 
     \caption[$M_*$ vs. $M_{HI}$]{Color contours show $M_*$ vs $M_{HI}$ of the full ALFALFA sample. 
      ALFALFA void dwarf (blue crosses) and stellar mass-matched wall dwarf (gold triangles) galaxies are overplotted. 
      We see that at a given stellar mass, void dwarf galaxies have higher \hi\ masses than wall dwarf galaxies. 
      This implies that void dwarf galaxies have lower SFEs than similar stellar mass galaxies in walls.}
     \label{fig:alfalfa_hi_vs_stell_mass}
\end{figure}
\begin{figure}[ht]
\centering
    \includegraphics[scale=0.5,trim = 0.7cm 1.0cm 2.0cm 1.9cm, clip=True]{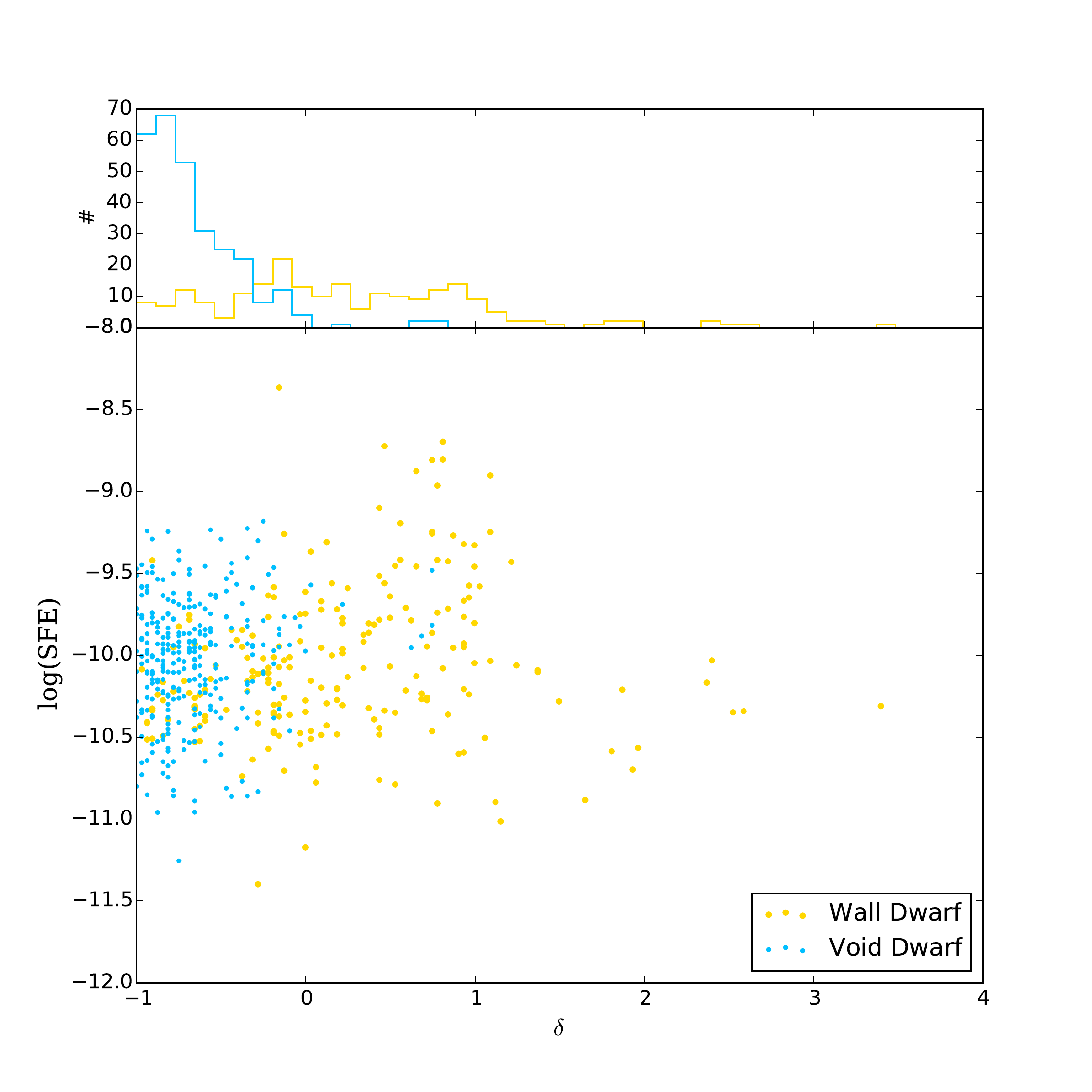} 
  \caption[SFE of ALFALFA dwarfs vs local density]{ 
    SFEs for void dwarf (blue) and stellar mass-matched 
    wall dwarf (gold) galaxies in the ALFALFA Dwarf sample as a function of local density. 
    We find that void galaxies are on average more isolated than wall galaxies. 
    Redshift space distributions show that \hi\ selected galaxies in ALFALFA   
    rarely appear in clusters, as seen in figure 5 of \cite{Moorman2014}.
  } 
 \label{fig:alfalfa_other}
\end{figure}

Figure \ref{fig:random_alfalfa_images} presents the distribution of SFEs as a function of $M_{HI}$, $M_*$, and SSFR. 
ALFALFA Dwarf void galaxies (blue crosses) and wall galaxies (gold triangles) are overplotted 
on the colored-contour of the full ALFALFA sample in each space. 
We see evidence of a population of $\sim10$ low-\hi\ mass dwarf galaxies with relatively high SFEs in walls. 
These 10 galaxies do not differ drastically from the other dwarf galaxies in stellar mass or SSFR. 
Figure \ref{fig:alfalfa_hi_vs_stell_mass} shows the $M_*$ vs $M_{HI}$ distribution of the ALFALFA dwarf galaxies. 
\cite{Huang2012} and \cite{Kreckel2012} show that ALFALFA galaxies are 
predominately \hi\ mass dominated below $\log(M_*)\sim9.5$ 
(falling below the dashed line in the figure) and stellar mass dominated at larger $M_*$. 
The high SFE wall galaxies fall below the $M_{HI}=M_*$ line, 
indicating that they have lower gas mass fractions than typical ALFALFA galaxies of similar stellar mass. 
We note that the void and wall dwarf galaxies plotted in 
Figure~\ref{fig:alfalfa_hi_vs_stell_mass} are stellar mass-matched, so 
the shift towards lower $M_{HI}$ at a given $M_*$ is important. 
Galaxies with low gas fractions typically show the following traits: 
low SFRs, low SSFRs, high SFEs, high metallicities, high extinctions, redder, and more evolved. 
These particular galaxies do have lower a SFR and SSFR, and a higher SFE, 
but they appear to be just as blue as the rest of the ALFALFA dwarf population 
and half of them are late-type while the other half are early-type as judged by 
the galaxies inverse concentration index. 
These galaxies live in relatively high local density regions as determined by 
the KDE estimator discussed in Section \ref{sec:methI} (see Figure \ref{fig:alfalfa_other} for SFEs as a function of local density). 

Measured on both time scales, the SFE distribution of void dwarf galaxies 
appears broader than that of wall dwarf galaxies. 
The broadened distribution towards lower SFEs could be indicative 
of \emph{marginally} lower SFEs in void dwarf galaxies. 
Furthermore, we see that the SFE distributions for void and wall galaxies both 
shift towards lower SFEs on shorter timescales, but we find no evidence of a stronger 
shift in either the void or wall samples.  
Perhaps this means the void environment does provide a sheltered life for dwarf galaxies, 
allowing them to retain their \hi\ gas allowing for fueling over longer periods.

\section{COMPARISON TO PREVIOUS RESULTS} 
\label{sec:compare}
Our results are consistent with previous studies regarding the variation of galaxy SSFRs with large-scale environment  
that indicate that galaxies in underdense environments have higher SSFRs than those in denser regions.    
\cite{Rojas2005} measure the SSFR of galaxies from an earlier data release of the SDSS and 
find that at fixed luminosity, void galaxies have higher SSFRs than wall galaxies. 
They also find that SFRs are very similar or slightly lower for void galaxies than wall galaxies, 
but this is expected, because void galaxies are generally less massive and SFRs are correlated with galaxy mass. 
Similarly, using a sample of faint galaxies from the 2dFGRS, \cite{vonBenda-Beckmann2008} find stronger star formation 
suppression in the field than in voids via the color-SSFR relation. 
That is, they find that galaxy colors are redder in denser regions 
than in voids. Because SSFR and color are strongly correlated, 
these authors make the claim that SSFRs must be lower in denser regions. 

At first glance, the recent results of \cite{Ricciardelli2014} appear to contradict our results, although this is likely 
a consquence of how the authors define ``void'' in their work. 
\cite{Ricciardelli2014} estimate the SSFR of ``void'' and ``shell'' galaxies within the SDSS DR7.  
These authors find that the average SSFR does \emph{not} vary with large-scale structure. 
These results do not necessarily contradict our results, because the definitions of 
LSS used in that work vary from what we define in this work.
The ``void'' galaxy sample used in \cite{Ricciardelli2014} is a sample of galaxies that live only in the void centers. 
That is, the authors use a void catalog that only includes the ``maximal sphere'' of each void. 
Assuming voids are non-spherical bodies comprised of multiple spheres to create an ellipsoidal shape, 
the ``maximal sphere'' of a void, 
is simply the sphere with the maximum radius within a single void. 
This allows the authors to use only galaxies in the void centers; 
however, they are comparing their void sample to a ``shell'' sample that likely contains void galaxies from the outskirts of the voids. 
\cite{Pan2012} show that the density profiles of voids are flat out to the edges, 
therefore properties of galaxies should not vary with void centric distance. 
The ``shells'' used in this paper are spherical shells from the maximal sphere radius, $R_{void}$, to $1.5R_{void}$. 
If all voids were perfectly spherical, this would work well to distinguish galaxies in underdense versus dense regions, 
but \cite{PanPhD2011} shows that voids in the nearby Universe 
are more ellipsoidal than spherical, with a tendency towards bring prolate.  
By accounting for only the maximal spheres of a void, 
these authors are using a sample littered with both void and wall galaxies as the ``shell'' sample. 
Comparing properties of void galaxies with a population that likely has a 
significant number of void galaxies mixed in, they would not notice a difference. 
It is clear from their distributions that they are lacking a large portion of quiescent galaxies that are present in our sample.   
Consistent with our findings, these authors find that the proportion of star forming galaxies in voids is higher and 
the fraction of passive galaxies are higher in denser regions. 

The SSFR trends seen in our work also agree with the high-resolution void simulation of \cite{Cen2011}, 
who predicts that void galaxies will have higher SSFRs than galaxies in denser regions. 
\cite{Cen2011} also finds a trend of increasing SSFRs with decreasing stellar mass. 
As we move into the dwarf regime, we find that galaxies with lower $M_*$ have higher SSFRs, 
with most (90\% void dwarf galaxies and 85\% of wall dwarf galaxies) 
having SSFRs high enough to double the stellar mass within a Hubble time, according to the $FUV$ method.

Our results on the SSFRs of ALFALFA galaxies are also in agreement with 
the results of \cite{Kreckel2012} regarding the environmental dependence of SSFR in an \hi\ selected sample. 
\cite{Kreckel2012} use the Void Galaxy Survey (VGS), a targeted \hi\ imaging survey, to map the \hi\ in 
galaxies in known voids in the nearby Universe. 
They compare 60 void galaxies from the VGS to a control sample of galaxies in 
average environments in a similar stellar mass range from the GALEX Arecibo SDSS Survey (GASS) catalog. 
They find that the SSFRs of \hi\ selected void galaxies are similar to SSFRs of galaxies in denser regions. 
While we find higher SSFRs in void galaxies for our NSA sample, the void and stellar mass-matched wall SSFRs 
of ALFALFA galaxies are very similar. 
\cite{Beygu2015} also use the VGS and find no difference in SSFR between VGS void galaxies 
and galaxies of similar stellar mass in denser environments 
(JCMT field galaxies, LV isolated and field galaxies, and ALFALFA Virgo cluster galaxies).  
The Galex Arecibo SDSS Survey (GASS) Data Release 3 \citep{Catinella2013} compares the 
SSFRs of a 760 galaxy sample across environments and finds that star formation is 
quenched at the group environments. The \hi\ sensitivity of this survey is higher than 
that of ALFALFA, however the GASS survey is designed to be complete at stellar 
masses above $10^{10} M_{\sun}$, a regime which includes only the highest decade 
of objects that we analyze.

Most works to date investigating the environmental dependence of star formation efficiencies 
have focused on clusters or isolated galaxies rather than on large-scale void environments. 
One work that does focus on the LSS dependence of SFEs is that of \cite{Kreckel2012}, who find 
only two galaxies within the VGS with $M_*>10^{10}M_{\sun}$ which they could 
compare to the GASS galaxies in denser regions. 
These two galaxies have higher SFEs ($10^{-8.6}$yr$^{-1}$ and $10^{-9.1}$yr$^{-1}$) than 
$M_*>10^{10}M_{\sun}$ GASS galaxies which had an average 
SFE of $10^{-9.5}$yr$^{-1}$ in this stellar mass range. 
\cite{Kreckel2012} further compare the VGS SFEs to a volume-limited sample of 447 ALFLFA galaxies 
($\log(M_{HI})>9$ within $0.007<z<0.024$). 
The 447 ALFALFA control galaxies live in average density regions ($1<\delta+1<10$) and are 
crossmatched to SDSS DR7 using 
cross-matching techniques found in \cite{Toribio2011Control}. 
The authors find hints (via a weak trend) that the VGS galaxies display higher SFEs 
than the volume-limited ALFALFA subset, but do not suggest that these results are significant. 
Our results for the full ALFALFA sample do not indicate that 
void galaxies, on the whole, are more effective at forming stars with their gas. 
This is in agreement with 
the analysis of \cite{Kreckel2011} who analyze the void simulation 
of \cite{Cen2011} and find that SFEs in voids are similar to those in 
denser regions across all magnitudes except in the dwarf galaxy regime. 
Further investigation of the VGS void galaxies by \cite{Beygu2015} reveals 
no difference in SFE between VGS void galaxies  
and galaxies of similar stellar mass in denser regions (JCMT field galaxies, 
LV isolated and field galaxies, and ALFALFA Virgo cluster galaxies).

Investigating the ALFALFA Dwarf SFE distributions, we find 
that the low-SFE end appears to have relatively more void galaxies than wall galaxies. 
This could be indicative of lower SFEs in voids, but is not statistically significant. 
If it is the case that void galaxies have lower SFEs, this would be in agreement with the work of \cite{Bigiel2010b} 
who find that the \hi\ column is correlated with SFE and is dependent on local environment. 
It may be the case that even small-scale groups of galaxies within voids provide a more isolated 
environment (i.e. lower $\delta$) than even the most isolated of galaxies outside of voids. 
See Figure~\ref{fig:alfalfa_other} which shows that 
void dwarf galaxies are significantly more isolated than wall dwarf galaxies.

\section{CONCLUSIONS}
\label{sec:sfr_conc}
We examine the effects of both large-scale and local environment on star formation properties of these two samples, 
focusing particularly on the faintest galaxies ($M_r\ge-17$).  
We utilize the NASA Sloan Atlas (NSA) 
which offers a set of cross-matched objects within the SDSS DR8-GALEX-ALFALFA footprint. 
From the NSA sample, we extract 113,145 galaxies with optical and UV information, 
and 8070 galaxies with information from all three catalogs. 
We determine the large-scale environment in which each galaxy lives using 
the void catalog of \cite{Pan2012} splitting the galaxies into ``void galaxies'' and ``wall galaxies.'' 
We match the stellar mass distributions of void and wall galaxies 
(thus controlling for the strong shift toward smaller galaxy mass in voids) and find the following results.

\begin{itemize}
\item Void dwarf ($M_r\ge-17$) galaxies have higher SSFRs than wall dwarf galaxies. 
	The trend  towards higher SSFRs in voids holds true on star formation 
	timescales of both $\sim10$ and $\sim100$ Myr, using SFR estimates 
	from \Ha\ lines and $FUV$ photometry, respectively. 
	We reproduce the trend towards higher SSFRs for bright void galaxies and 
	reveal that the shift towards higher SSFRs in cosmic voids extends down to magnitudes as faint as $M_r=-13$.  
\item When we limit our sample to galaxies containing enough \hi\ to be detected by ALFALFA, 
	we remove the passive galaxy population. After stellar mass matching the void and wall ALFALFA galaxies, 
	we notice no difference in the SSFR distributions. This further supports the finding that 
	ALFALFA detects primarily blue, star forming galaxies. 
\item Comparing only the active star forming sequences in the NSA and ALFALFA catalogs shows a 
	shift towards lower SSFRs in ALFALFA. It is possible that \hi\ selected galaxies are more 
	dust-reddened on average than optically selected star forming galaxies. 
	This result is consistent with the finding in \cite{Moorman2015} 
	that the blue cloud in ALFALFA is somewhat redder than the blue cloud of 
	the optically selected sample.  Figure 1 in \cite{Moorman2015} shows a similar shift of 
	the blue cloud towards redder colors than the blue cloud galaxy distribution of the optically selected sample.
\item Investigating the dwarf galaxy population within the NSA and ALFALFA samples, we find that dwarf galaxies 
	have higher SSFRs and lower SFEs than typical brighter galaxies. 
	This indicates that dwarf galaxies are likely to be actively forming stars at a relatively high rate. 
	The lower SFEs could also imply that dwarf galaxies are able to retain their gas more easily than brighter galaxies 
	(lower SFEs are caused by large \hi\ masses rather than low SFRs in the ratio SFE=SFR/$M_{HI}$).  
\item We do not find statistical evidence that the void environment has an impact 
	on the SFEs of dwarf galaxies.  
	This is unsurprising, given that galaxies in our void and wall samples have  
	strong \hi\ emission and similar stellar mass distributions. 
	Assuming that the two samples {\it are} drawn from the same distribution, it would appear that 
	the wall dwarf sample is lacking galaxies around log(SFE)$\sim-10$. 
	Dwarf galaxies have been forming stars at relatively high rates in recent times (10Myr), 
	but due to survey sensitivities, we can only see dwarf galaxies in the very nearby Universe. 
	Because it is difficult to identify true voids in the nearby volume, 
	we are unable to make a statistically significant comparison of the 
	star formation efficiencies of void dwarf and wall dwarf galaxies.

 \end{itemize}

\section*{Acknowledgments} 
The authors would like to acknowledge the work of the entire
ALFALFA collaboration team in observing, flagging, and extracting 
the catalog of galaxies used in this work. 
The ALFALFA team at Cornell is supported by NSF grants AST-0607007 
and AST-1107390 to RG and MPH and by grants from the Brinson Foundation.
This study was supported by NSF grant AST-1410525 at Drexel University. 
We thank the referee for their helpful comments.

The NASA-Sloan Atlas was created by Michael Blanton, with extensive help and testing from Eyal Kazin, 
Guangtun Zhu, Adrian Price-Whelan, John Moustakas, Demitri Muna, Renbin Yan and Benjamin Weaver. 
Renbin Yan provided the detailed spectroscopic measurements for each SDSS spectrum. 
David Schiminovich kindly provided the input GALEX images. 
We thank Nikhil Padmanabhan, David Hogg, Doug Finkbeiner and David Schlegel 
for their work on SDSS image infrastructure.

We acknowledge the support of NASA, the Centre National d'Etudes Spatiales (CNES) of France, 
and the Korean Ministry of Science and Technology for the 
construction, operation, and data analysis of the GALEX mission.

Funding for the creation and distribution of the SDSS Archive has been
provided by the Alfred P. Sloan Foundation, the Participating
Institutions, the National Aeronautics and Space Administration, the
National Science Foundation, the U.S. Department of Energy, the
Japanese Monbukagakusho, and the Max Planck Society. The SDSS Web site
is http://www.sdss.org/.

The SDSS is managed by the Astrophysical Research Consortium (ARC) for
the Participating Institutions. The Participating Institutions are The
University of Chicago, Fermilab, the Institute for Advanced Study, the
Japan Participation Group, The Johns Hopkins University, the Korean
Scientist Group, Los Alamos National Laboratory, the
Max-Planck-Institute for Astronomy (MPIA), the Max-Planck-Institute
for Astrophysics (MPA), New Mexico State University, University of
Pittsburgh, Princeton University, the United States Naval Observatory,
and the University of Washington.

\bibliographystyle{./apj}
\bibliography{./bibliography3}

\begin{thebibliography}{}
\expandafter\ifx\csname natexlab\endcsname\relax\def\natexlab#1{#1}\fi

\bibitem[{{Aihara} {et~al.}(2011){Aihara}, {Allende Prieto}, {An}, {Anderson},
  {Aubourg}, {Balbinot}, {Beers}, {Berlind}, {Bickerton}, {Bizyaev}, {Blanton},
  {Bochanski}, {Bolton}, {Bovy}, {Brandt}, {Brinkmann}, {Brown}, {Brownstein},
  {Busca}, {Campbell}, {Carr}, {Chen}, {Chiappini}, {Comparat}, {Connolly},
  {Cortes}, {Croft}, {Cuesta}, {da Costa}, {Davenport}, {Dawson}, {Dhital},
  {Ealet}, {Ebelke}, {Edmondson}, {Eisenstein}, {Escoffier}, {Esposito},
  {Evans}, {Fan}, {Femen{\'{\i}}a Castell{\'a}}, {Font-Ribera}, {Frinchaboy},
  {Ge}, {Gillespie}, {Gilmore}, {Gonz{\'a}lez Hern{\'a}ndez}, {Gott}, {Gould},
  {Grebel}, {Gunn}, {Hamilton}, {Harding}, {Harris}, {Hawley}, {Hearty}, {Ho},
  {Hogg}, {Holtzman}, {Honscheid}, {Inada}, {Ivans}, {Jiang}, {Johnson},
  {Jordan}, {Jordan}, {Kazin}, {Kirkby}, {Klaene}, {Knapp}, {Kneib},
  {Kochanek}, {Koesterke}, {Kollmeier}, {Kron}, {Lampeitl}, {Lang}, {Le Goff},
  {Lee}, {Lin}, {Long}, {Loomis}, {Lucatello}, {Lundgren}, {Lupton}, {Ma},
  {MacDonald}, {Mahadevan}, {Maia}, {Makler}, {Malanushenko}, {Malanushenko},
  {Mandelbaum}, {Maraston}, {Margala}, {Masters}, {McBride}, {McGehee},
  {McGreer}, {M{\'e}nard}, {Miralda-Escud{\'e}}, {Morrison}, {Mullally},
  {Muna}, {Munn}, {Murayama}, {Myers}, {Naugle}, {Neto}, {Nguyen}, {Nichol},
  {O'Connell}, {Ogando}, {Olmstead}, {Oravetz}, {Padmanabhan},
  {Palanque-Delabrouille}, {Pan}, {Pandey}, {P{\^a}ris}, {Percival},
  {Petitjean}, {Pfaffenberger}, {Pforr}, {Phleps}, {Pichon}, {Pieri}, {Prada},
  {Price-Whelan}, {Raddick}, {Ramos}, {Reyl{\'e}}, {Rich}, {Richards}, {Rix},
  {Robin}, {Rocha-Pinto}, {Rockosi}, {Roe}, {Rollinde}, {Ross}, {Ross},
  {Rossetto}, {S{\'a}nchez}, {Sayres}, {Schlegel}, {Schlesinger}, {Schmidt},
  {Schneider}, {Sheldon}, {Shu}, {Simmerer}, {Simmons}, {Sivarani}, {Snedden},
  {Sobeck}, {Steinmetz}, {Strauss}, {Szalay}, {Tanaka}, {Thakar}, {Thomas},
  {Tinker}, {Tofflemire}, {Tojeiro}, {Tremonti}, {Vandenberg}, {Vargas
  Maga{\~n}a}, {Verde}, {Vogt}, {Wake}, {Wang}, {Weaver}, {Weinberg}, {White},
  {White}, {Yanny}, {Yasuda}, {Yeche}, \& {Zehavi}}]{Aihara2011}
{Aihara}, H., {Allende Prieto}, C., {An}, D., {et~al.} 2011, \apjs, 195, 26

\bibitem[{{Alpaslan} {et~al.}(2015){Alpaslan}, {Driver}, {Robotham},
  {Obreschkow}, {Andrae}, {Cluver}, {Kelvin}, {Lange}, {Owers}, {Taylor},
  {Andrews}, {Bamford}, {Bland-Hawthorn}, {Brough}, {Brown}, {Colless},
  {Davies}, {Eardley}, {Grootes}, {Hopkins}, {Kennedy}, {Liske}, {Lara-Lopez},
  {Lopez-Sanchez}, {Loveday}, {Madore}, {Mahajan}, {Meyer}, {Moffett},
  {Norberg}, {Penny}, {Pimbblet}, {Popescu}, {Seibert}, \&
  {Tuffs}}]{Alpaslan2015}
{Alpaslan}, M., {Driver}, S., {Robotham}, A.~S.~G., {et~al.} 2015, ArXiv
  e-prints, arXiv:1505.05518

\bibitem[{{Benson} {et~al.}(2002){Benson}, {Lacey}, {Baugh}, {Cole}, \&
  {Frenk}}]{Benson2002}
{Benson}, A.~J., {Lacey}, C., {Baugh}, C., {Cole}, S., \& {Frenk}, C. 2002, in
  Astronomical Society of the Pacific Conference Series, Vol. 254,
  Extragalactic Gas at Low Redshift, ed. J.~S. {Mulchaey} \& J.~T. {Stocke},
  354

\bibitem[{{Beygu} {et~al.}(2015){Beygu}, {Kreckel}, {van der Hulst},
  {Peletier}, {Jarrett}, {van de Weygaert}, {van Gorkom}, \&
  {Arag{\'o}n-Calvo}}]{Beygu2015}
{Beygu}, B., {Kreckel}, K., {van der Hulst}, J.~M., {et~al.} 2015, ArXiv
  e-prints, arXiv:1501.02577

\bibitem[{{Bigiel} {et~al.}(2010){Bigiel}, {Leroy}, {Walter}, {Blitz},
  {Brinks}, {de Blok}, \& {Madore}}]{Bigiel2010b}
{Bigiel}, F., {Leroy}, A., {Walter}, F., {et~al.} 2010, \aj, 140, 1194

\bibitem[{{Blanton} {et~al.}(2011){Blanton}, {Kazin}, {Muna}, {Weaver}, \&
  {Price-Whelan}}]{Blanton2011}
{Blanton}, M.~R., {Kazin}, E., {Muna}, D., {Weaver}, B.~A., \& {Price-Whelan},
  A. 2011, \aj, 142, 31

\bibitem[{{Blanton} {et~al.}(2003){Blanton}, {Lin}, {Lupton}, {Maley}, {Young},
  {Zehavi}, \& {Loveday}}]{Blanton2003SpectraSDSS}
{Blanton}, M.~R., {Lin}, H., {Lupton}, R.~H., {et~al.} 2003, \aj, 125, 2276

\bibitem[{{Blanton} \& {Roweis}(2007)}]{Blanton_kcorrect}
{Blanton}, M.~R., \& {Roweis}, S. 2007, \aj, 133, 734

\bibitem[{{Bruzual} \& {Charlot}(2003)}]{Bruzual2003}
{Bruzual}, G., \& {Charlot}, S. 2003, \mnras, 344, 1000

\bibitem[{{Catinella} {et~al.}(2010){Catinella}, {Schiminovich}, {Kauffmann},
  {Fabello}, {Wang}, {Hummels}, {Lemonias}, {Moran}, {Wu}, {Giovanelli},
  {Haynes}, {Heckman}, {Basu-Zych}, {Blanton}, {Brinchmann}, {Budav{\'a}ri},
  {Gon{\c c}alves}, {Johnson}, {Kennicutt}, {Madore}, {Martin}, {Rich},
  {Tacconi}, {Thilker}, {Wild}, \& {Wyder}}]{Catinella2010}
{Catinella}, B., {Schiminovich}, D., {Kauffmann}, G., {et~al.} 2010, \mnras,
  403, 683

\bibitem[{{Catinella} {et~al.}(2013){Catinella}, {Schiminovich}, {Cortese},
  {Fabello}, {Hummels}, {Moran}, {Lemonias}, {Cooper}, {Wu}, {Heckman}, \&
  {Wang}}]{Catinella2013}
{Catinella}, B., {Schiminovich}, D., {Cortese}, L., {et~al.} 2013, \mnras, 436,
  34

\bibitem[{{Cen}(2011)}]{Cen2011}
{Cen}, R. 2011, \apj, 741, 99

\bibitem[{{Choi} {et~al.}(2010){Choi}, {Han}, \& {Kim}}]{Choi2010}
{Choi}, Y.-Y., {Han}, D.-H., \& {Kim}, S.~S. 2010, Journal of Korean
  Astronomical Society, 43, 191

\bibitem[{{da Costa} {et~al.}(1988){da Costa}, {Pellegrini}, {Sargent},
  {Tonry}, {Davis}, {Meiksin}, {Latham}, {Menzies}, \& {Coulson}}]{daCosta1988}
{da Costa}, L.~N., {Pellegrini}, P.~S., {Sargent}, W.~L.~W., {et~al.} 1988,
  \apj, 327, 544

\bibitem[{{El-Ad} \& {Piran}(1997)}]{ElAdPiran1997}
{El-Ad}, H., \& {Piran}, T. 1997, \apj, 491, 421

\bibitem[{{Elbaz} {et~al.}(2007){Elbaz}, {Daddi}, {Le Borgne}, {Dickinson},
  {Alexander}, {Chary}, {Starck}, {Brandt}, {Kitzbichler}, {MacDonald},
  {Nonino}, {Popesso}, {Stern}, \& {Vanzella}}]{Elbaz2007}
{Elbaz}, D., {Daddi}, E., {Le Borgne}, D., {et~al.} 2007, \aap, 468, 33

\bibitem[{{Fukugita} {et~al.}(1996){Fukugita}, {Ichikawa}, {Gunn}, {Doi},
  {Shimasaku}, \& {Schneider}}]{Fukugita1996}
{Fukugita}, M., {Ichikawa}, T., {Gunn}, J.~E., {et~al.} 1996, \aj, 111, 1748

\bibitem[{{Geller} \& {Huchra}(1989)}]{GellerHuchra1989}
{Geller}, M.~J., \& {Huchra}, J.~P. 1989, Science, 246, 897

\bibitem[{{Giovanelli} {et~al.}(2005{\natexlab{a}}){Giovanelli}, {Haynes},
  {Kent}, {Perillat}, {Saintonge}, {Brosch}, {Catinella}, {Hoffman},
  {Stierwalt}, {Spekkens}, {Lerner}, {Masters}, {Momjian}, {Rosenberg},
  {Springob}, {Boselli}, {Charmandaris}, {Darling}, {Davies}, {Garcia Lambas},
  {Gavazzi}, {Giovanardi}, {Hardy}, {Hunt}, {Iovino}, {Karachentsev},
  {Karachentseva}, {Koopmann}, {Marinoni}, {Minchin}, {Muller}, {Putman},
  {Pantoja}, {Salzer}, {Scodeggio}, {Skillman}, {Solanes}, {Valotto}, {van
  Driel}, \& {van Zee}}]{ALFALFAI}
{Giovanelli}, R., {Haynes}, M.~P., {Kent}, B.~R., {et~al.} 2005{\natexlab{a}},
  \aj, 130, 2598

\bibitem[{{Giovanelli} {et~al.}(2005{\natexlab{b}}){Giovanelli}, {Haynes},
  {Kent}, {Perillat}, {Catinella}, {Hoffman}, {Momjian}, {Rosenberg},
  {Saintonge}, {Spekkens}, {Stierwalt}, {Brosch}, {Masters}, {Springob},
  {Karachentsev}, {Karachentseva}, {Koopmann}, {Muller}, {van Driel}, \& {van
  Zee}}]{ALFALFAII}
---. 2005{\natexlab{b}}, \aj, 130, 2613

\bibitem[{{Goldberg} {et~al.}(2005){Goldberg}, {Jones}, {Hoyle}, {Rojas},
  {Vogeley}, \& {Blanton}}]{Goldberg2005}
{Goldberg}, D.~M., {Jones}, T.~D., {Hoyle}, F., {et~al.} 2005, \apj, 621, 643

\bibitem[{{Goldberg} \& {Vogeley}(2004)}]{Goldberg2004}
{Goldberg}, D.~M., \& {Vogeley}, M.~S. 2004, \apj, 605, 1

\bibitem[{{Gunn} {et~al.}(1998){Gunn}, {Carr}, {Rockosi}, {Sekiguchi}, {Berry},
  {Elms}, {de Haas}, {Ivezi{\'c}}, {Knapp}, {Lupton}, {Pauls}, {Simcoe},
  {Hirsch}, {Sanford}, {Wang}, {York}, {Harris}, {Annis}, {Bartozek},
  {Boroski}, {Bakken}, {Haldeman}, {Kent}, {Holm}, {Holmgren}, {Petravick},
  {Prosapio}, {Rechenmacher}, {Doi}, {Fukugita}, {Shimasaku}, {Okada}, {Hull},
  {Siegmund}, {Mannery}, {Blouke}, {Heidtman}, {Schneider}, {Lucinio}, \&
  {Brinkman}}]{Gunn1998}
{Gunn}, J.~E., {Carr}, M., {Rockosi}, C., {et~al.} 1998, \aj, 116, 3040

\bibitem[{{Haynes} {et~al.}(2011){Haynes}, {Giovanelli}, {Martin}, {Hess},
  {Saintonge}, {Adams}, {Hallenbeck}, {Hoffman}, {Huang}, {Kent}, {Koopmann},
  {Papastergis}, {Stierwalt}, {Balonek}, {Craig}, {Higdon}, {Kornreich},
  {Miller}, {O'Donoghue}, {Olowin}, {Rosenberg}, {Spekkens}, {Troischt}, \&
  {Wilcots}}]{Haynes2011}
{Haynes}, M.~P., {Giovanelli}, R., {Martin}, A.~M., {et~al.} 2011, \aj, 142,
  170

\bibitem[{{Hoeft} {et~al.}(2006){Hoeft}, {Yepes}, {Gottl{\"o}ber}, \&
  {Springel}}]{Hoeft2006}
{Hoeft}, M., {Yepes}, G., {Gottl{\"o}ber}, S., \& {Springel}, V. 2006, \mnras,
  371, 401

\bibitem[{{Hopkins} {et~al.}(2003){Hopkins}, {Miller}, {Nichol}, {Connolly},
  {Bernardi}, {G{\'o}mez}, {Goto}, {Tremonti}, {Brinkmann}, {Ivezi{\'c}}, \&
  {Lamb}}]{Hopkins2003}
{Hopkins}, A.~M., {Miller}, C.~J., {Nichol}, R.~C., {et~al.} 2003, \apj, 599,
  971

\bibitem[{{Hoyle} {et~al.}(2005){Hoyle}, {Rojas}, {Vogeley}, \&
  {Brinkmann}}]{Hoyle2005}
{Hoyle}, F., {Rojas}, R.~R., {Vogeley}, M.~S., \& {Brinkmann}, J. 2005, \apj,
  620, 618

\bibitem[{{Hoyle} \& {Vogeley}(2002)}]{HoyleVogeley2002}
{Hoyle}, F., \& {Vogeley}, M.~S. 2002, \apj, 566, 641

\bibitem[{{Hoyle} {et~al.}(2012){Hoyle}, {Vogeley}, \& {Pan}}]{Hoyle2012}
{Hoyle}, F., {Vogeley}, M.~S., \& {Pan}, D. 2012, \mnras, 426, 3041

\bibitem[{{Huang} {et~al.}(2012{\natexlab{a}}){Huang}, {Haynes}, {Giovanelli},
  \& {Brinchmann}}]{Huang2012}
{Huang}, S., {Haynes}, M.~P., {Giovanelli}, R., \& {Brinchmann}, J.
  2012{\natexlab{a}}, \apj, 756, 113

\bibitem[{{Huang} {et~al.}(2012{\natexlab{b}}){Huang}, {Haynes}, {Giovanelli},
  {Brinchmann}, {Stierwalt}, \& {Neff}}]{Huang2012Dwarf}
{Huang}, S., {Haynes}, M.~P., {Giovanelli}, R., {et~al.} 2012{\natexlab{b}},
  \aj, 143, 133

\bibitem[{{Icke}(1984)}]{Icke1984}
{Icke}, V. 1984, \mnras, 206, 1P

\bibitem[{{Karachentsev} {et~al.}(2004){Karachentsev}, {Karachentseva},
  {Huchtmeier}, \& {Makarov}}]{Karachentsev2004}
{Karachentsev}, I.~D., {Karachentseva}, V.~E., {Huchtmeier}, W.~K., \&
  {Makarov}, D.~I. 2004, \aj, 127, 2031

\bibitem[{{Kennicutt}(1998)}]{Kennicutt1998}
{Kennicutt}, Jr., R.~C. 1998, \araa, 36, 189

\bibitem[{{Koposov} {et~al.}(2009){Koposov}, {Yoo}, {Rix}, {Weinberg},
  {Macci{\`o}}, \& {Escud{\'e}}}]{Koposov2009}
{Koposov}, S.~E., {Yoo}, J., {Rix}, H.-W., {et~al.} 2009, \apj, 696, 2179

\bibitem[{{Kreckel} {et~al.}(2011){Kreckel}, {Joung}, \& {Cen}}]{Kreckel2011}
{Kreckel}, K., {Joung}, M.~R., \& {Cen}, R. 2011, \apj, 735, 132

\bibitem[{{Kreckel} {et~al.}(2012){Kreckel}, {Platen}, {Arag{\'o}n-Calvo}, {van
  Gorkom}, {van de Weygaert}, {van der Hulst}, \& {Beygu}}]{Kreckel2012}
{Kreckel}, K., {Platen}, E., {Arag{\'o}n-Calvo}, M.~A., {et~al.} 2012, \aj,
  144, 16

\bibitem[{{Lee} {et~al.}(2009){Lee}, {Gil de Paz}, {Tremonti}, {Kennicutt},
  {Salim}, {Bothwell}, {Calzetti}, {Dalcanton}, {Dale}, {Engelbracht}, {Funes},
  {Johnson}, {Sakai}, {Skillman}, {van Zee}, {Walter}, \& {Weisz}}]{Lee2009}
{Lee}, J.~C., {Gil de Paz}, A., {Tremonti}, C., {et~al.} 2009, \apj, 706, 599

\bibitem[{{Lupton} {et~al.}(2001){Lupton}, {Gunn}, {Ivezi{\'c}}, {Knapp}, \&
  {Kent}}]{Lupton2001}
{Lupton}, R., {Gunn}, J.~E., {Ivezi{\'c}}, Z., {Knapp}, G.~R., \& {Kent}, S.
  2001, in Astronomical Society of the Pacific Conference Series, Vol. 238,
  Astronomical Data Analysis Software and Systems X, ed. F.~R. {Harnden}, Jr.,
  F.~A. {Primini}, \& H.~E. {Payne}, 269

\bibitem[{{Martin} {et~al.}(2005){Martin}, {Fanson}, {Schiminovich},
  {Morrissey}, {Friedman}, {Barlow}, {Conrow}, {Grange}, {Jelinsky},
  {Milliard}, {Siegmund}, {Bianchi}, {Byun}, {Donas}, {Forster}, {Heckman},
  {Lee}, {Madore}, {Malina}, {Neff}, {Rich}, {Small}, {Surber}, {Szalay},
  {Welsh}, \& {Wyder}}]{GALEX2005}
{Martin}, D.~C., {Fanson}, J., {Schiminovich}, D., {et~al.} 2005, \apjl, 619,
  L1

\bibitem[{{Moorman} {et~al.}(2014){Moorman}, {Vogeley}, {Hoyle}, {Pan},
  {Haynes}, \& {Giovanelli}}]{Moorman2014}
{Moorman}, C.~M., {Vogeley}, M.~S., {Hoyle}, F., {et~al.} 2014, \mnras, 444,
  3559

\bibitem[{{Moorman} {et~al.}(2015){Moorman}, {Vogeley}, {Hoyle}, {Pan},
  {Haynes}, \& {Giovanelli}}]{Moorman2015}
---. 2015, \apj, 810, 108

\bibitem[{{Pan}(2011)}]{PanPhD2011}
{Pan}, D.~C. 2011, PhD thesis, Drexel University

\bibitem[{{Pan} {et~al.}(2012){Pan}, {Vogeley}, {Hoyle}, {Choi}, \&
  {Park}}]{Pan2012}
{Pan}, D.~C., {Vogeley}, M.~S., {Hoyle}, F., {Choi}, Y.-Y., \& {Park}, C. 2012,
  \mnras, 421, 926

\bibitem[{{Ricciardelli} {et~al.}(2014){Ricciardelli}, {Cava}, {Varela}, \&
  {Quilis}}]{Ricciardelli2014}
{Ricciardelli}, E., {Cava}, A., {Varela}, J., \& {Quilis}, V. 2014, \mnras,
  445, 4045

\bibitem[{{Rojas} {et~al.}(2005){Rojas}, {Vogeley}, {Hoyle}, \&
  {Brinkmann}}]{Rojas2005}
{Rojas}, R.~R., {Vogeley}, M.~S., {Hoyle}, F., \& {Brinkmann}, J. 2005, \apj,
  624, 571

\bibitem[{{Salim} {et~al.}(2007){Salim}, {Rich}, {Charlot}, {Brinchmann},
  {Johnson}, {Schiminovich}, {Seibert}, {Mallery}, {Heckman}, {Forster},
  {Friedman}, {Martin}, {Morrissey}, {Neff}, {Small}, {Wyder}, {Bianchi},
  {Donas}, {Lee}, {Madore}, {Milliard}, {Szalay}, {Welsh}, \& {Yi}}]{Salim2007}
{Salim}, S., {Rich}, R.~M., {Charlot}, S., {et~al.} 2007, \apjs, 173, 267

\bibitem[{{Stanonik} {et~al.}(2009){Stanonik}, {Platen}, {Arag{\'o}n-Calvo},
  {van Gorkom}, {van de Weygaert}, {van der Hulst}, \&
  {Peebles}}]{Stanonik2009}
{Stanonik}, K., {Platen}, E., {Arag{\'o}n-Calvo}, M.~A., {et~al.} 2009, \apjl,
  696, L6

\bibitem[{{Strauss} {et~al.}(2002){Strauss}, {Weinberg}, {Lupton}, {Narayanan},
  {Annis}, {Bernardi}, {Blanton}, {Burles}, {Connolly}, {Dalcanton}, {Doi},
  {Eisenstein}, {Frieman}, {Fukugita}, {Gunn}, {Ivezi{\'c}}, {Kent}, {Kim},
  {Knapp}, {Kron}, {Munn}, {Newberg}, {Nichol}, {Okamura}, {Quinn}, {Richmond},
  {Schlegel}, {Shimasaku}, {SubbaRao}, {Szalay}, {Vanden Berk}, {Vogeley},
  {Yanny}, {Yasuda}, {York}, \& {Zehavi}}]{Strauss2002}
{Strauss}, M.~A., {Weinberg}, D.~H., {Lupton}, R.~H., {et~al.} 2002, \aj, 124,
  1810

\bibitem[{{Toribio} {et~al.}(2011){Toribio}, {Solanes}, {Giovanelli}, {Haynes},
  \& {Masters}}]{Toribio2011Control}
{Toribio}, M.~C., {Solanes}, J.~M., {Giovanelli}, R., {Haynes}, M.~P., \&
  {Masters}, K.~L. 2011, \apj, 732, 92

\bibitem[{{van de Weygaert} \& {van Kampen}(1993)}]{vandeWeygaert1993}
{van de Weygaert}, R., \& {van Kampen}, E. 1993, \mnras, 263, 481

\bibitem[{{von Benda-Beckmann} \& {M{\"u}ller}(2008)}]{vonBenda-Beckmann2008}
{von Benda-Beckmann}, A.~M., \& {M{\"u}ller}, V. 2008, \mnras, 384, 1189

\bibitem[{{Willick} {et~al.}(1997){Willick}, {Courteau}, {Faber}, {Burstein},
  {Dekel}, \& {Strauss}}]{Willick1997}
{Willick}, J.~A., {Courteau}, S., {Faber}, S.~M., {et~al.} 1997, \apjs, 109,
  333

\bibitem[{{Yan}(2011)}]{Yan2011}
{Yan}, R. 2011, \aj, 142, 153

\bibitem[{{Yan} \& {Blanton}(2012)}]{Yan2012}
{Yan}, R., \& {Blanton}, M.~R. 2012, \apj, 747, 61

\bibitem[{{York} {et~al.}(2000){York}, {Adelman}, {Anderson}, {Anderson},
  {Annis}, {Bahcall}, {Bakken}, {Barkhouser}, {Bastian}, {Berman}, {Boroski},
  {Bracker}, {Briegel}, {Briggs}, {Brinkmann}, {Brunner}, {Burles}, {Carey},
  {Carr}, {Castander}, {Chen}, {Colestock}, {Connolly}, {Crocker}, {Csabai},
  {Czarapata}, {Davis}, {Doi}, {Dombeck}, {Eisenstein}, {Ellman}, {Elms},
  {Evans}, {Fan}, {Federwitz}, {Fiscelli}, {Friedman}, {Frieman}, {Fukugita},
  {Gillespie}, {Gunn}, {Gurbani}, {de Haas}, {Haldeman}, {Harris}, {Hayes},
  {Heckman}, {Hennessy}, {Hindsley}, {Holm}, {Holmgren}, {Huang}, {Hull},
  {Husby}, {Ichikawa}, {Ichikawa}, {Ivezi{\'c}}, {Kent}, {Kim}, {Kinney},
  {Klaene}, {Kleinman}, {Kleinman}, {Knapp}, {Korienek}, {Kron}, {Kunszt},
  {Lamb}, {Lee}, {Leger}, {Limmongkol}, {Lindenmeyer}, {Long}, {Loomis},
  {Loveday}, {Lucinio}, {Lupton}, {MacKinnon}, {Mannery}, {Mantsch}, {Margon},
  {McGehee}, {McKay}, {Meiksin}, {Merelli}, {Monet}, {Munn}, {Narayanan},
  {Nash}, {Neilsen}, {Neswold}, {Newberg}, {Nichol}, {Nicinski}, {Nonino},
  {Okada}, {Okamura}, {Ostriker}, {Owen}, {Pauls}, {Peoples}, {Peterson},
  {Petravick}, {Pier}, {Pope}, {Pordes}, {Prosapio}, {Rechenmacher}, {Quinn},
  {Richards}, {Richmond}, {Rivetta}, {Rockosi}, {Ruthmansdorfer}, {Sandford},
  {Schlegel}, {Schneider}, {Sekiguchi}, {Sergey}, {Shimasaku}, {Siegmund},
  {Smee}, {Smith}, {Snedden}, {Stone}, {Stoughton}, {Strauss}, {Stubbs},
  {SubbaRao}, {Szalay}, {Szapudi}, {Szokoly}, {Thakar}, {Tremonti}, {Tucker},
  {Uomoto}, {Vanden Berk}, {Vogeley}, {Waddell}, {Wang}, {Watanabe},
  {Weinberg}, {Yanny}, {Yasuda}, \& {SDSS Collaboration}}]{York2000}
{York}, D.~G., {Adelman}, J., {Anderson}, Jr., J.~E., {et~al.} 2000, \aj, 120,
  1579

\end{thebibliography}
\end{document}